\documentclass[sigconf]{acmart}

\usepackage{url}
\usepackage{hyperref}
\usepackage{amsfonts}
\usepackage[T1]{fontenc}
\usepackage{color}
\usepackage{enumitem}
\usepackage{multirow}
\usepackage{float}
\usepackage{subcaption}
\usepackage{appendix}
\usepackage{tabularx}
\usepackage{longtable}

\newcommand{\xhdr}[1]{\vspace{0.2mm}\noindent{{\bf #1.}}}

\copyrightyear{2022}
\acmYear{2022}
\setcopyright{rightsretained}
\acmConference[CHI '22]{CHI Conference on Human Factors in Computing Systems}{April 29-May 5, 2022}{New Orleans, LA, USA}
\acmBooktitle{CHI Conference on Human Factors in Computing Systems (CHI '22), April 29-May 5, 2022, New Orleans, LA, USA}
\acmDOI{10.1145/3491102.3517527}
\acmISBN{978-1-4503-9157-3/22/04}

\begin{document}

\title[Does Terminology Affect Human Perceptions and Evaluations of ADM Systems?]{``Look! It's a Computer Program! It's an Algorithm! It's AI!'': Does Terminology Affect Human Perceptions and Evaluations of Algorithmic Decision-Making Systems?}

\author{Markus Langer}
\affiliation{
    \institution{Universität des Saarlandes}
    \country{Germany}}
\email{markus.langer@uni-saarland.de}

\author{Tim Hunsicker}
\affiliation{
    \institution{Universität des Saarlandes}
    \country{Germany}}
\email{tim.hunsicker@uni-saarland.de}

\author{Tina Feldkamp}
\affiliation{
    \institution{Universität des Saarlandes}
    \country{Germany}}
\email{tina.feldkamp@uni-saarland.de}

\author{Cornelius J. König}
\affiliation{
    \institution{Universität des Saarlandes}
    \country{Germany}}
\email{ckoenig@mx.uni-saarland.de}

\author{Nina Grgi\'{c}-Hla\v{c}a}
\affiliation{
    \institution{Max Planck Institute for Software Systems, Max Planck Institute for Research on Collective Goods}
    \country{Germany}}
\email{nghlaca@mpi-sws.org}

\renewcommand{\shortauthors}{Langer et al.}


\begin{abstract}

In the media, in policy-making, but also in research articles, algorithmic decision-making (ADM) systems are referred to as algorithms, artificial intelligence, and computer programs, amongst other terms. We hypothesize that such terminological differences can affect people's perceptions of properties of ADM systems, people's evaluations of systems in application contexts, and the replicability of research as findings may be influenced by terminological differences. In two studies (\textit{N} = 397, \textit{N} = 622), we show that terminology does indeed affect laypeople's perceptions of system properties (e.g., perceived complexity) and evaluations of systems (e.g., trust). Our findings highlight the need to be mindful when choosing terms to describe ADM systems, because terminology can have unintended consequences, and may impact the robustness and replicability of HCI research. Additionally, our findings indicate that terminology can be used strategically (e.g., in communication about ADM systems) to influence people's perceptions and evaluations of these systems.

\end{abstract}



\maketitle


\section{Introduction} \label{sec:intro}
 
When the public discusses algorithmic decision-making systems (ADM systems) -- systems that either automate decision-making or support human decision-making -- when journalists report about such systems, and when policy-makers develop regulations about such systems, there is a variety of terms used to refer to them. For instance, newspaper articles refer to such systems as intelligent systems \cite{livecareer2018}, as algorithms \cite{oren2016howto}, or robotic systems \cite{ertz2021eight}. Likewise, there is large variety in terminology used to refer to ADM systems in policy-making documents. For instance, within the European Commission's ``Ethics Guidelines for Trustworthy AI'' \cite{euai2021}, the authors refer to ADM systems as algorithms, artificial intelligence, AI technologies, AI systems, and robots whereas the General Data Protection Regulation (GDPR) refers to ADM systems as automated means. 

Similar variation in the terminology used to refer to ADM systems also occurs in research investigating interactions between humans and ADM systems. In such research, researchers develop materials where they describe the respective system to their participants. For instance, researchers might be interested in how trustworthy their participants perceive a system to be \cite{lee2018understanding} or may investigate whether participants accept the respective system \cite{langer2020anybody}. In such studies, research has used the terms algorithm \cite{lee2018understanding}, automated system \cite{keel2018feasibility}, artificial intelligence \cite{lee2021included}, computer program \cite{grgic2018human}, machine learning \cite{gonzalez2019s}, sophisticated statistical model \cite{dietvorst2015algorithm}, or robot \cite{otting2018importance} -- all to refer to a system that either automates decision-making or that supports human decision-making in a variety of application contexts (e.g., for systems that support hiring decisions \cite{langer2020anybody}, medical decisions \cite{keel2018feasibility}, or bail decisions at court \cite{grgic2018human}). 

Whereas all those terms reflect a similar idea -- a system that interacts with humans -- they might induce very different mental pictures, expectations, and thoughts associated with the ADM system in question. More generally, presenting participants a system using the term “automated system” versus “algorithm” versus “artificial intelligence” may affect how people perceive and evaluate these systems. On the one hand, this may affect the robustness and replicability of HCI research as findings may vary between studies only because of terminological differences. For instance, people's acceptance of an ADM system in medicine might differ depending on whether the system is described as an ``algorithm'' or as a ``computer program''. On the other hand, communicating about ADM systems (e.g., in policy-making) using the term ``automated system'' versus ``artificial intelligence'' might alter what people expect when they hear the respective term. For instance, an ``automated system'' might sound less advanced compared to using ``artificial intelligence'' and this could affect initial perceptions of trustworthiness with respect to the system in question because ``artificial intelligence'' is associated with a system with more potential than an ``automated system''.

In this paper, we propose that terminology crucially affects the ways in which people perceive and evaluate ADM systems. More precisely, we argue that the choice of the term used to refer to ADM systems will affect people's perceptions about the properties of the system (e.g., perceived complexity) as well as people's evaluation of the system (e.g., trust evaluations) in application contexts. We conducted two experimental, between-subject studies to test whether terminology matters, and if different terminology can cause different effects in communication about ADM systems. In the first study, we varied ten terms that research has used to refer to ADM systems to explore how this affects people’s perceptions of properties of the respective systems. Additionally, we examined terminological effects on people's evaluation of whether systems or humans are better able to conduct a set of different tasks (e.g., medical diagnoses, criminal recidivism prediction). In the second study, we used vignettes of a well-known study in HCI by \citet{lee2018understanding} and varied the term used within those vignettes to test if evaluations of fairness and trust in application contexts differ depending on the terminology used to refer to ADM systems. 

\xhdr{Contributions}
In this paper, we contribute to research on HCI by showing that terminological differences affect
\begin{itemize}
 \item Human perceptions of properties of ADM systems (e.g., perceived complexity)
 \item Human evaluations of systems (e.g., trust)
\end{itemize}

We thereby highlight the importance of terminology in communication about ADM systems. On the one hand, variation in terminology can have unintended negative effects on the robustness and replicability of HCI research. On the other hand, terminology can be used strategically to steer human perceptions and evaluations of such systems.


\section{Related Work} \label{sec:related_work}
\subsection{Why terminology may matter} \label{sec:terminology_matters}

Studies throughout disciplines have shown the importance of terminology as it can affect human perceptions, emotions, and behavior \cite{eitzel2017citizen, puhl2020words, shank2020software}. We propose that terminological differences used to refer to ADM systems in HCI will affect people's perceptions and evaluations of ADM systems. Specifically, research has used a variety of terms to refer to ADM systems \cite{hou2021expert,langer2021future, shank2021humans} which applies to the description of ADM systems within papers but more crucially to the communication about ADM systems when presenting them to research participants. For instance, \citet{wang2020factors} told their participants that an ``algorithm'' processes their MTurk work history, decides who will get a promotion (i.e., become a master worker), and then asked participants to evaluate fairness of the algorithm-based decision. In a school admission scenario, \citet{marcinkowski2020implications} told participants that an ``AI technology'' analyzes applicant data and recommends applicant approval or rejection. They also asked for participants' evaluations of the fairness of the AI technology's decision. In a work scheduling setting, \citet{uhde2020fairness} told participants that a ``system'' decides who gets vacation and asked them to report how they perceive and evaluate system-based decisions for scheduling. Even in single papers presenting multiple studies, terminology to refer to ADM systems might vary. For example, \citet{longoni2019resistance} present multiple studies on the acceptance of AI in healthcare (e.g., in skin cancer screening). In their studies they described to participants that the respective ADM system is a ``computer [that] uses an algorithm'', ``a computer that is capable of artificial intelligence'', ``a computer program'' or ``a well-trained algorithm'' that provides outputs that help to make medical decisions. As another example, \citet{binns2018s} asked participants about their evaluations of situations where a ``computer system'' or a ``predictive model'' is used to decide whether a person should receive a promotion.

Terminology effects might be especially influential in previous studies because participants often received limited information regarding the system in question. In fact, \citet{langer2021future} reviewed research on people's perceptions and evaluations of automated systems in different decision-making situations (e.g., management, healthcare). In many of the studies they reviewed, the term to refer to the system was the main experimental manipulation as it was this term that informed people about the fact that there is an ADM system automating decisions or supporting decision-making. For instance, \citet{nagtegaal2021impact} told participants that a ``computer, using an automated algorithm'', decides about travel reimbursement or evaluates employee performance. In \citet{langer2020anybody}, the only information their participants, who had to record responses to job interview questions, received was that a ``computer will automatically analyze the audio recordings and evaluate [their] answers''. In both these examples, the focus seems to be on the automation of a decision by an ADM system without further specifying this system. In further examples, \citet{shaffer2013patients} described to participants who had to rate the expertise of doctors that a ``doctor [...] indicates she is going to use a decision aid [computer program]'' and \citet{dietvorst2015algorithm} described to their participants who had the option to use outputs by a model as additional information to forecast student performance that ``the admissions office had created a statistical model that was designed to forecast student performance'' and provide the additional information that this model is ``sophisticated''. In both these examples, ADM systems were introduced to support decision-making but there was no further information about underlying technology or about, for instance, how the system produces its outputs. In other work (e.g. \cite{otting2018importance}), terminology such as robot may have been chosen deliberately to describe an embodied ADM system and to additionally anthropomorphize the system by describing it as a humanoid robot. Importantly, in these and in many more studies investigating people's reactions to (partly) automated decision-making \cite{langer2021future}, there was limited additional information regarding the functionalities or performance of the system, limited information regarding how the system works, and especially a limited rationale regarding why respective authors chose a specific term to describe an ADM system to participants. Without additional information about how a respective system works, or about functionalities of a system, people need to rely on salient aspects within study information to form their perceptions and evaluations of the respective situation \cite{aguinis2014best, atzmuller2010experimental}. This kind of salient information can be the term used to refer to ADM systems. 

To decide which terms to investigate, we drew on Langer and Landers' \cite{langer2021future} review that provides an overview on the terms research has used to describe ADM systems. Additionally, we added two more terms that have been used to refer to ADM systems in studies not included in Langer and Landers' review. We added the term ``technical system'' \cite{montague2009empirically} as a term that is very generic, as well as the term ``sophisticated statistical model'' \cite{dietvorst2015algorithm} as a term that is very specific. Table \ref{tab:terms_used} presents the final set of terms we decided to investigate as well as sample sources that have used these terms in their studies.

\begin{table}[ht]
  \centering
  \caption{Terminological differences to refer to ADM systems with the 10 terms used in Study 1 and exemplary studies that have used these terms.}
  \label{tab:terms_used}
  \begin{tabular}{l|l}
  \hline
  \textbf{Term} & \textbf{Exemplary Study} \\
  \hline
  Algorithm & \citet{lee2018understanding} \\
  Automated system & \citet{keel2018feasibility} \\
  Artificial intelligence & \citet{marcinkowski2020implications} \\
  Computer & \citet{langer2020anybody} \\
  Computer program & \citet{grgic2019human} \\
  Decision support system & \citet{shibl2013factors} \\
  Machine learning & \citet{gonzalez2019s} \\
  Technical system & \citet{montague2009empirically} \\
  Robot & \citet{otting2018importance} \\
  Sophisticated statistical model & \citet{dietvorst2015algorithm} \\
  \hline
  \end{tabular}
  \end{table}

\subsection{Consequences of terminological differences}

In this paper, we empirically investigate two broad consequences of terminological differences when referencing ADM systems. First, we explore consequences for perceptions of properties of ADM systems. For this, we investigate what kind of properties people associate with different references to ADM systems, irrespective of the context in which the ADM system is used. In other words, to shed light on the properties associated with the respective term to describe ADM systems, we chose to only vary the term and to not give any additional information (e.g., on system functionalities or the application context). Understanding how terminology affects perceptions of properties associated with the entity is important as this might provide us with insights regarding what basic properties are associated with different terms, which might allow conclusions regarding more downstream consequences (e.g., acceptance of systems).

Second we explore consequences for evaluations of ADM systems in application contexts. This means we explore whether different terminology to describe an ADM system in an application context can differently affect people's evaluations of the respective system. This is important because it allows insights regarding whether and to what extent using different terms to describe ADM systems in application contexts may affect people's evaluations of ADM systems (e.g., regarding trust, fairness).

\subsubsection{Consequences of terminological differences for perceptions of the properties of ADM systems}

We chose to assess six properties associated with ADM systems: tangibility, complexity, controllability, familiarity, anthropomorphism, and machine competence. We chose these properties because they can be evaluated without putting ADM systems in an application context and because research has shown them to be related to more downstream consequences such as acceptance of systems, or human behavior in the interaction with systems.

\textbf{Tangibility.} Tangibility is associated with people having a shape in mind when they think about a term and whether a term is associated with an entity humans can touch \cite{glikson2020human}. People may interact differently with agents having a physical appearance compared to disembodied agents, perceive them as more socially present \cite{lee2006physically, li2015benefit}, and may have different expectations regarding relationship building with more tangible entities \cite{glikson2020human}. With respect to the terms we use, we imagine that terms such as ``computer'' or ``robot'' are more likely perceived as tangible compared to terms such as ``algorithm'' or ``artificial intelligence'' since the former have a shape while the latter reflect disembodied manifestations of ADM systems.

\textbf{Complexity.} In this paper, high complexity would mean people believe the entity described by the term is hard to understand, including its functionalities and its design-process, and for which it is hard to comprehend how it works \cite{gell2002complexity, nagtegaal2021impact}. Perceived complexity can be associated with the acceptance of systems \cite{nagtegaal2021impact} and with beliefs about system quality \cite{elsbach2019new}. Regarding the terms, ``computer program'' might be perceived to be less complex compared to ``artificial intelligence'' because even though people might not understand how computer programs work, artificial intelligence may be associated with more complex technologies.

\textbf{Controllability.} Controllability is associated with whether people believe humans can control the behavior of the entity described by the term. Perceived controllability relates to the acceptance of systems \cite{venkatesh2003user, verbert2013visualizing}. With respect to the different terms, ``computer'' might be associated with an entity that is more controllable compared to ``robot'' because people have already operated the former and might believe that the latter is acting more autonomously \cite{o2020human}. 

\textbf{Familiarity.} If a term is associated with something that is familiar, people have already heard of the term, have had experience with using the entity associated with this term, and believe that the entity is something that is a part of everyday life. Familiarity is, for instance, associated with better acceptance of systems \cite{venkatesh2003user, de2015sharing}. We, for instance, imagine that ``computer program'' is perceived to be more familiar than ``machine learning'' since computer programs are something people use every day, whereas machine learning reflects a more specific concept where only experts would say that it is familiar to them.

\textbf{Anthropomorphism.} Anthropomorphism refers to whether people perceive the term describing an entity as possessing human-like characteristics \cite{de2016almost, epley2007seeing}. For instance, anthropomorphism can be associated with believing that an entity has intentions or makes autonomous decisions. Anthropomorphism is an important variable in agent design where virtual agents can be designed more or less anthropomorphic in order to affect human-agent interaction patterns \cite{culley2013note, araujo2018living, de2016almost}. Regarding different terms, it is possible that people perceive ``technical system'' to be less associated with human-like characteristics compared to ``robot'' or ``artificial intelligence'' that are often presented as having or evolving these characteristics in popular media.

\textbf{Machine competence.} Under machine competence, we understand whether a term is associated with an entity that has great capabilities and strong potential regarding its successful application in different contexts \cite{glikson2020human}. Machine competence is usually associated with high expectations regarding the performance of ADM systems and may thus determine whether people use a respective system \cite{glikson2020human, hoff2015trust, lee2004trust}. Regarding the different terms, the capabilities that people ascribe to ``artificial intelligence'' might be stronger compared to capabilities associated with ``decision support system'' because artificial intelligence may sound like something with broader application possibilities than decision support systems. 

Considering the different perceptions different terminologies regarding ADM systems can invoke, we propose the following research question:

\textbf{Research Question 1:} Does varying the terminology regarding ADM systems affect people’s perceptions of the properties of ADM systems?~\footnote{Before data collection for the respective studies started, we preregistered the research questions, dependent variables that we wanted to capture, experimental manipulations, data exclusion plan, data analysis plan, and planned number of participants to include in the studies. The respective blinded preregistrations are available under https://aspredicted.org/LDC\_GSM and https://aspredicted.org/NTE\_WND}

\subsubsection{Consequences for evaluations of ADM systems in application contexts}

Up to this point, we have focused on perceptions or properties associated with ADM systems without considering the application context in which these systems may operate. Consequences of terminological differences become even more important when considering the evaluation of systems in specific application contexts. Specifically, varying the term used to refer to ADM systems may affect whether people positively or negatively evaluate the use of said system in a respective context, and may lead to a lack of acceptance or disuse just due to terminological differences and not actual differences in system-design or functionalities \cite{jacovi2021formalizing}. 

To investigate whether terminological differences affect people's evaluations of ADM systems in application contexts we a) examine whether terminological differences affect evaluations regarding the ability of systems to conduct a set of different tasks, and b) investigate whether terminological differences affect evaluations of fairness and trust in systems as well as robustness and replicability of research by replicating a well-known study on evaluations of ADM systems in application contexts (i.e., \cite{lee2018understanding}).

Regarding a), we thus chose a set of different tasks that are associated with the use of ADM systems (e.g., making shopping recommendations, evaluating applicant documents, providing therapy recommendations in medicine) to explore whether the term used to refer to ADM systems affects whether people evaluate a system to be able to perform a respective task. We chose a set of tasks that reflects a variety of application contexts as well as different tasks in single application contexts (e.g., in medicine). Since recent work shows emerging interest in understanding the tasks where people believe systems to perform better or at least equally well as human beings (see e.g., \cite{castelo2019task, dietvorst2020people, langer2021future, lee2018understanding}), we wanted to investigate whether the evaluation of the performance of systems in such tasks also depends on the terminology to describe the system.

\textbf{Research Question 2:} Does varying the term to refer to ADM systems affect people’s evaluation regarding the performance of systems in various tasks?

Regarding b), instead of devising a novel study paradigm, we chose to replicate Lee's \cite{lee2018understanding} well-known study on evaluations of fairness and trust in different application contexts and varied the terminology she used to refer to the respective ADM system described in her study. She presented participants with textual vignettes that described one of four application contexts (work assignment, work scheduling, hiring, and work evaluation) where an ADM system described with the term ``algorithm'' provided decisions that affect human decision-recipients. She found that for tasks that afford human skills (hiring, work evaluation) people evaluated the algorithm to be less fair and participants trusted the algorithm less in these application contexts compared to human decisions. In contrast, she found less, and non-significant, differences between the human manager and the algorithm for tasks that afford mechanical skills (work assignment, work scheduling).

We propose that using a different term than ``algorithm'' might affect the results of her study and consequently the conclusions we can draw from the study. Specifically, instead of ``algorithm'', it is equally possible to refer to the system that produces a decision as an ``automated system'' which may affect people's evaluations of the respective system. For example, if people evaluate algorithms to be more capable of conducting a specific task compared to automated systems, this could lead to different levels of trust. Similarly, if people evaluate automated systems to be more consistent in decision-making than algorithms, this could affect fairness perceptions. If we find that terminological differences indeed affect the conclusions we draw from the respective study (e.g., for certain terms we find stronger, significant effects, whereas for others we find smaller, non-significant ones), we might need to infer that parts of variability in findings from previous research were due to differing terminology to refer to ADM systems \cite{langer2021future}. Additionally, finding that terminological differences can affect the conclusions we draw from research would indicate that it is necessary to be more mindful when choosing the terminology to describe ADM systems to participants in studies.

In addition to the evaluation of trust and fairness that \citet{lee2018understanding} investigated in her study, we chose to also capture perceived procedural justice \cite{colquitt2001dimensionality} as a related concept. Furthermore, since \citet{lee2018understanding} investigated different application contexts in her study, we took the opportunity to investigate whether terminological differences affect evaluations of systems differently depending on the application context. If this would be the case, we would find an interaction effect in our study results, indicating that the effect of different terminology may also depend on the task for which a respective ADM system is used. In other words, terminological effects may be stronger for one task than for another. Overall, we thus propose the following research questions:

\textbf{Research Question 3:} Does varying the term to refer to ADM systems affect people's evaluation of ADM systems (in our case evaluations of trust, fairness, and procedural justice)?

\textbf{Research Question 4:} Will the term used to refer to ADM systems and the task for which ADM systems is used interact to affect evaluations of ADM systems?


\section{Methodology} \label{sec:methodology}

\subsection{Study 1}

Study 1 investigated people's perceptions of properties associated with different terms to refer to ADM systems. Additionally, Study 1 shed light on people's perceptions regarding whether the term to refer to the system affects the perceived ability of systems to perform several different tasks. 

\subsubsection{Sample}
\begin{figure}
  \centering
  \includegraphics[width=1.1\columnwidth]{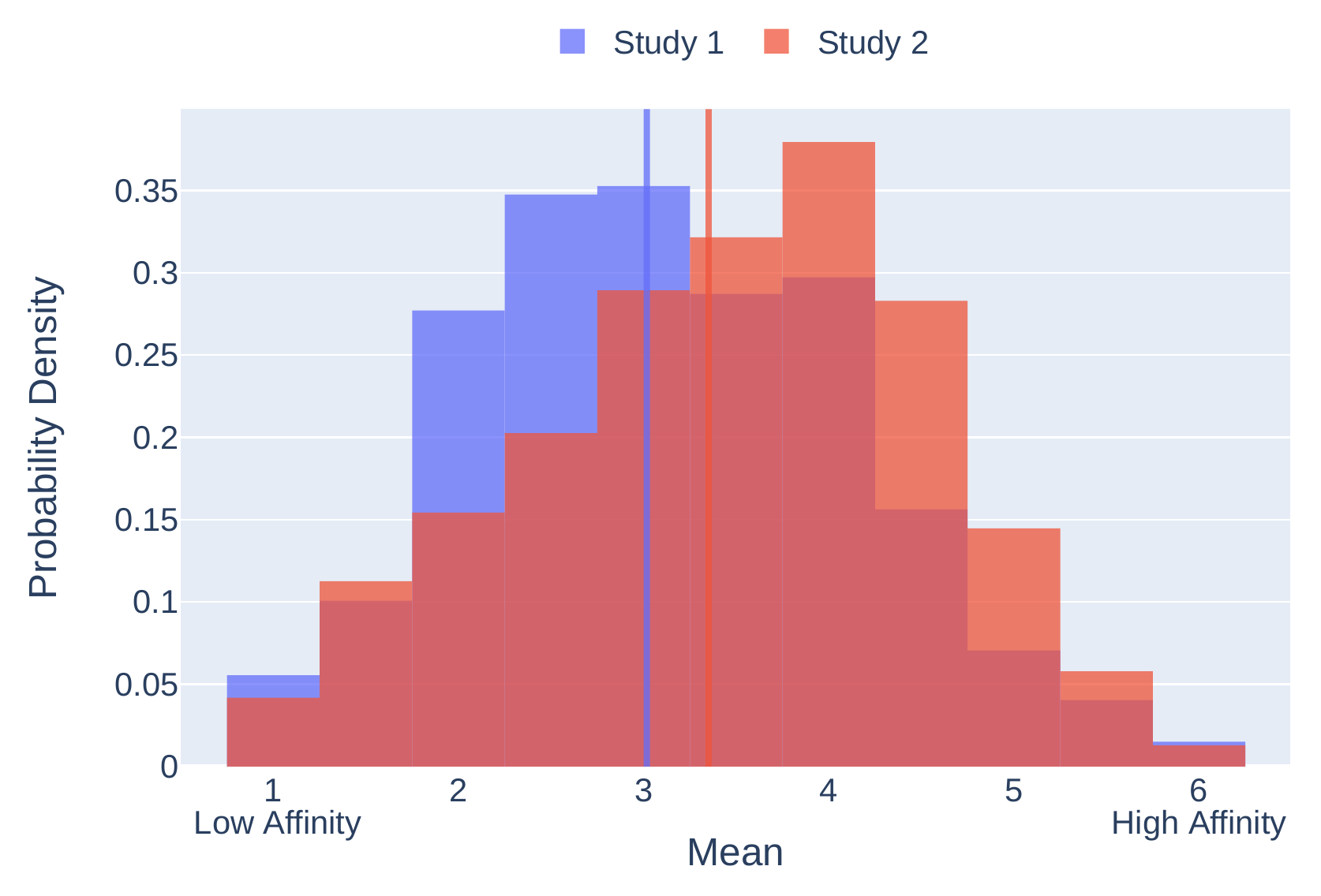}
  \caption{Distribution of participants' responses to the affinity for technology scale in Study 1 (blue) and Study 2 (red), and the corresponding mean values.}
  \label{fig:affinity_distribution}
  \Description{This figure shows the distribution of participants' responses to the affinity for technology scale in Study 1 and Study 2. The left bold line shows the mean value for Study 1 and the right line shows the mean value for Study 2 which was slightly higher. Also this figure shows the distribution of participant responses indicating that they are similar in both studies.}
\end{figure}

We gathered data on Prolific. The only inclusion criterion was that participants were native English-speakers and 18 years or older. Completing the study took on average 8 (\textit{SD} = 2) minutes and participants received 1.27 British Pounds as payment. We gathered data from \textit{N} = 417 participants. We excluded 6 participants because they did not recall the correct term that was used in their version of the study as well as 14 participants because they failed the included attention check item. The final sample consisted of \textit{N} = 397 participants (65\% female; 35\% male), with a mean age of 35 years (\textit{SD} = 13); 69\% of participants were employed. 21\% of participants were students. Furthermore, 7\% of participants reported their highest education level as ``attended high school'', 19\% reported that they have a high school degree, 18\% reported a 2-year community/technical/professional/trade college degree, 35\% reported a 4-year college or university degree, and 20\% reported a graduate degree or PhD. A majority of participants was from the United Kingdom (78\%), with the rest of participants coming from South Africa (5\%), Australia (4\%), the US (3\%), Ireland (2\%), and small numbers of participants from various other countries. Since we imagined that participants' interest in and prior exposure to different technologies may affect how they perceive the different terms, we measured participants' affinity for technology \cite{franke2019personal} as a possible control variable. The mean value of participants' affinity for technology was \textit{M} = 3.02 (\textit{SD} = 1.03) (See Figure \ref{fig:affinity_distribution}). This indicates a mean value that would correspond to the label ``slightly disagree'' in the response options to the affinity for technology scale but also some between-participant variation regarding affinity for technology.

\subsubsection{Procedure}
Study 1 was conducted online and followed a randomized experimental design with 10 between-subject conditions. This means that a single participant was presented with exactly one of the terms to refer to ADM systems presented in Table \ref{tab:terms_used}. Following, each time we use ``the term'' we use it as a placeholder for the experimentally manipulated terms.

Participants accessed the study through a link that directed them to the first page in the online questionnaire tool (we used SoSci Survey). After providing informed consent, participants were randomly assigned to their respective experimental condition. First, participants received a set of items that asked for their perceptions of properties associated with the entity described by the respective term (see a screenshot in Appendix Figure \ref{fig:appendix_study1_perceptions}). Participants responded to items assessing tangibility, complexity, controllability, familiarity, anthropomorphism, and machine competence. Second, participants were asked to evaluate how well they believe ADM systems will perform different tasks in comparison to humans. Specifically, we used thirteen tasks commonly associated with the use of systems that covered a range of different application settings (see Section \ref{sec:measures1}; see a screenshot in Appendix Figure \ref{fig:appendix_study1_betterthanhuman}). Third, participants responded to a scale assessing their affinity for technology \cite{franke2019personal} which served as a control measure to investigate whether general affinity for technology affects participants' perceptions of respective terms. Fourth, participants reported demographic information (gender, age, education level, whether they are students, and whether they are employed).~\footnote{Participants also responded to the negative attitudes towards robot scale \cite{nomura2006altered}, where we replaced the term robot with the respective term to refer to ADM systems. Participants also responded to the Godspeed scale \cite{bartneck2009measurement}. Furthermore, they were asked to report ``What is `the term' for you?'', and we included an item asking for their knowledge regarding the respective term. Finally, participants were asked to respond to the question ``Could you give us an example of `the term' that you have heard of or already used for work or in your free time?'' Results for these measures can be made available upon request.} 

To ensure data quality, participants responded to two attention check items: the first one asked them to respond ``strongly disagree'' to the respective attention check item, the second one asked them to report which of the ten terms they were presented with during the study. 

\subsubsection{Measures} \label{sec:measures1}
Unless otherwise stated, participants responded to the items measuring the dependent variables on a scale from 1 (strongly disagree) to 5 (strongly agree) -- ``the term'' was replaced with one of the terms reflecting our experimental manipulation. All items for Study 1 can be found in the Appendix (see Appendix Table \ref{tab:items_study1}). As a measure of scale reliability, for all scales with more than two items, we report Cronbach's $\alpha$; for two item scales, we report the Spearman-Brown correlation as suggested by \citet{eisinga2013reliability}.

Tangibility was measured with two self-developed items and assessed whether people have a clear picture or shape in mind when thinking about the respective term \cite{glikson2020human}. A sample item was ``When I think of `the term', I have a clear picture in mind'' (Spearman-Brown correlation = .67).

Complexity was measured with three self-developed items and assessed whether people believe that the term reflects something complex and non-comprehensible \cite{elsbach2019new}. A sample item was ```the term' is complex'' (Cronbach's $\alpha$ = .70).~\footnote{In our survey, we had included five items to capture complexity which can be found in Appendix Table \ref{tab:items_study1}. Two of the items led to a low Cronbach's $\alpha$. Following \citet{allen2008coefficient} and \citet{peterson1994meta}, we removed these items from the scale. Note that exclusion or inclusion of these items did not substantially change our interpretations for complexity. In both cases, we would find that different terminology affects perceived complexity.}

Controllability was measured with two self-developed items that assessed whether people believe that the term reflects something that is controllable for humans or whether the term reflects something that acts autonomously \cite{o2020human}. A sample item was ```the term' is controllable by humans'' (Spearman-Brown correlation = 0.78).~\footnote{In our survey, we had included three items to capture controllability which can be found in Appendix Table \ref{tab:items_study1}. One of the items led to a low Cronbach's $\alpha$. Following \citet{allen2008coefficient} and \citet{peterson1994meta}, we removed this item from the scale. Note that exclusion or inclusion of this item did not substantially change our interpretations for controllability. In both cases, we would find that different terminology affects perceived controllability.}

Familiarity was measured with three self-developed items that should reflect familiarity as described by \citet{luhmann2000familiarity}. A sample item was ```the term' is something I encounter in everyday life'' (Cronbach's $\alpha$ = .68).

Anthropomorphism was measured with eight items taken from \citet{shank2018attributions}. A sample item was ```the term' has intentions'' (Cronbach's $\alpha$ = 0.82).

Machine competence was measured with six self-developed items which assessed perceptions of high capabilities associated with the term \cite{glikson2020human}. A sample item was ```the term' has great potential in terms of what it can be used for'' (Cronbach's $\alpha$ = 0.77).

Perceived ability to perform several tasks in comparison to humans was measured for thirteen tasks: shopping recommendations, evaluating applicant documents, scheduling work, predicting criminal recidivism, making medical diagnoses, evaluating X-rays and MRIs, predicting the weather, evaluating job interviews, therapy recommendations in medicine, diagnosing mental illness, identifying faces, assessing dangerous situations while driving, and predicting the spread of infectious diseases. For instance, participants read: ```the term' can make shopping recommendations'' and were then asked to rate this statement on a scale from 1 (worse than a human) to 5 (better than a human) with the middle category 3 (as good as a human). 

The control variable affinity for technology was measured with four items taken from \citet{franke2019personal}. For this measure, we used the original response scale from 1 (completely disagree) to 6 (completely agree) (Cronbach's $\alpha$ = .83).

\subsection{Study 2}
To investigate effects of terminological differences on evaluations of ADM systems (e.g., trust), and to explore whether terminological differences affect robustness and replicability of research, Study 2 followed the methodology of \citet{lee2018understanding} and thus partly replicated her study that examined human evaluations of ADM system-based decisions in different application scenarios. 

\subsubsection{Sample}
We again gathered data on Prolific. The inclusion criteria were that participants were native English-speakers and 18 years or older, and that they had participated in at least 10 studies on Prolific and had a 100\% approval rate. Completing the study took on average 4 minutes (\textit{SD} = 1) and participants received 0.67 British Pound as payment. We gathered data from \textit{N} = 722 participants. We excluded 24 participants because they did not recall the correct term that was used in their version of the study indicating that they were not attentive. Furthermore, we excluded 76 participants because they failed the included attention check item. The final sample consisted of \textit{N} = 622 participants (62\% female; 38\% male), with a mean age of 36 years (\textit{SD} = 13); 71\% of participants were employed. 17\% of participants self-reported to be students. Furthermore, 7\% of participants reported their highest education level as ``attended high school'', 20\% reported that they have a high school degree, 15\% reported a 2-year community/technical/professional/trade college degree, 42\% reported a 4-year college or university degree, and 16\% reported a graduate degree or PhD. A majority of participants was from the United Kingdom (77\%), with the rest of participants coming from South Africa (4\%), the US (4\%), Canada (3\%) Ireland (3\%), and small numbers of participants from various other countries. Interest in technology as well as prior exposure to technology could affect the evaluation of the terms in application context, we thus again measured participants' affinity for technology. The mean value of participants' affinity for technology was \textit{M} = 3.35 (\textit{SD} = 1.08) (See Figure \ref{fig:affinity_distribution}). Similar to Study 1, this indicates a mean value that would correspond to the label ``slightly disagree'' in the response options to the affinity for technology scale but some between-participant variation.

\subsubsection{Reducing the number of terms to include in Study 2} \label{sec:lessterms}

\begin{figure}
   \centering
   \includegraphics[width=\columnwidth]{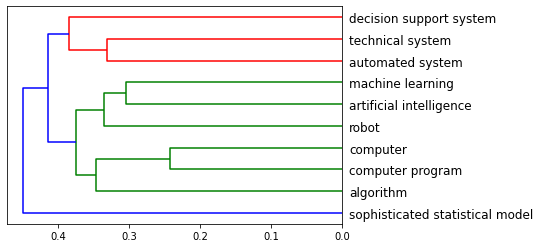}
   \caption{Clusters based on the semantic similarity between terms. The semantic similarity was estimated using Google's USE \cite{cer2018universal}.}
    \label{fig:semantic_clusters}
    \Description{This figure is fully described in the text.}
\end{figure}

To reduce the complexity of the study, we wanted to use fewer terms in Study 2. To determine which terms to keep, we used Google's Universal Sentence Encoder (USE) \cite{cer2018universal} to estimate the semantic similarity between the 10 terms used in Study 1. Specifically, we estimated which terms are semantically most similar and which ones more different, and used this information to determine which terms to include in Study 2. Google's USE has been trained on unsupervised training data from web sources such as Wikipedia and discussion forums, and supervised data from the Stanford Natural Language Inference corpus \cite{bowman2015large}, and was shown to perform well on the Semantic Textual Similarity Benchmark \cite{cer2017semeval}. We utilized USE to encode our terms into 512-dimensional embedding vectors and, as suggested by \citet{cer2018universal}, we calculated the angular similarity between these vectors in order to estimate the semantic similarity between the terms. We applied hierarchical clustering on the resulting distance matrix~\footnote{The distance between two terms encoded into 512-dimensional vectors \textbf{u} and \textbf{v} is calculated as $\text{dist}(\textbf{u}, \textbf{v}) = \text{arccos}\left(\frac{ \textbf{u} \cdot \textbf{v}}{\|\textbf{u}\| \|\textbf{v}\|}\right) / \pi$, where $\textbf{u} \cdot \textbf{v}$ is the dot product of \textbf{u} and \textbf{v}, and $\|\ast\|$ is the Euclidean norm of its argument $\ast$.}, using the UPGMA algorithm implemented in SciPy \cite{2020SciPy-NMeth}. The resulting clusters are shown in Figure \ref{fig:semantic_clusters}.
  
Based on the results for the semantic similarity analysis, we argue that there are four high-level clusters. The first cluster consisted of the term sophisticated statistical model. The second cluster included algorithm, computer program, and computer. The third cluster included robot, artificial intelligence, and machine learning. The fourth cluster included automated system, technical system, and decision support system. For Study 2, we decided to use one term from each cluster which is why we chose: sophisticated statistical model, computer program, artificial intelligence, and automated system. Furthermore, we included the term that \citet{lee2018understanding} also used in her study: algorithm. Since all of these terms may reflect disembodied manifestations of ADM systems, we decided to also include one of the terms that previous work has used to describe an embodied ADM system \cite{otting2018importance}: robot.

\subsubsection{Procedure}
Study 2 was conducted online and followed a randomized 7 (condition term: six different terms plus human condition) x 2 (condition task: work assignment versus work evaluation) experimental between-subject design. To be clear, each participant was thus presented with exactly one term in one task. As described in Section \ref{sec:lessterms}, we focused only on a subset of terms in comparison to Study 1: algorithm, automated system, artificial intelligence, computer program, robot, and sophisticated statistical model. In line with \citet{lee2018understanding}, we also included a human condition where participants read ``a manager'' instead of a term referring to a system. Our second experimental factor was the application context where we had the conditions work assignment and work evaluation. We decided to include these as they were also used in \citet{lee2018understanding} and because participants in Lee's study perceived system decisions to be much fairer and participants trusting these decisions to a larger extent for work assignment compared to work evaluation contexts. 

After providing informed consent, participants received initial information on the experimental setting. Specifically, participants read that ``In the situation below, `the term' makes a decision autonomously without human intervention.'' After this information participants were introduced to the respective decision situation reflecting their experimental condition. Specifically, we used the textual vignettes developed by \citet{lee2018understanding} verbatim with two changes. First, we replaced the term ``algorithm'' that she used in these vignettes in her study with the respective term of the given experimental condition. Additionally, we standardized the name of the person in the textual vignette to be ``Chris'' in every condition. The vignettes for the tasks were the following (see also screenshots in Appendix Figures \ref{fig:appendix_study2_workevaluation} and \ref{fig:appendix_study2_workassignment}):

\begin{itemize}
\item{Work assignment: ``In the following situation, `the term' makes a decision autonomously without human intervention. In a manufacturing factory, ‘the term’ assigns their employees to check and update certain components of the machinery to prevent any critical operation failures. The component assignment is based on data that show how often different components have worn out and broken down in the past. Chris works in the manufacturing factory. ‘The term’ assigns him to check a specific component of the machinery and he does the maintenance work on it.}
\item{Work evaluation: ``In the following situation, `the term' makes a decision autonomously without human intervention. In a customer service center, `the term'  evaluates employees by analyzing the content and tone of their calls with customers. Chris works at the customer service center. Based on past call recordings, `the term' evaluates his performance.''}
\end{itemize}

After reading their respective vignette, participants were asked to respond to the fairness and trust item as well as to the procedural justice items. Afterwards, participants responded to the affinity for technology items and to the demographic questions (gender, age, education level, whether they are studying, and whether they are employed).

Throughout Study 2, participants responded to two attention check items. The first one asked them to check the response option ``to a large extent''. The second one asked them to report which of the terms they were presented with during the study.

\subsubsection{Measures}
All items for Study 2 can be found in the Appendix (see Appendix Table \ref{tab:items_study2}). Fairness was measured with one item taken from \citet{lee2018understanding} that differed slightly with respect to the decision situation. In the case of work assignment this item was: ``How fair or unfair is it for Chris that `the term' assigns him to check a specific component of the machinery and he does the maintenance work on it?'' In the case of work evaluation this item was: ``How fair or unfair is it for Chris that `the term' evaluates his performance?'' Participants responded to this item on the same scale used by \citet{lee2018understanding} ranging from 1 (very unfair) to 7 (very fair).

Trust was measured with one item taken from \citet{lee2018understanding} that differed slightly with respect to the decision situation. In the case of work assignment this item was: ``How much do you trust that `the term' makes a good-quality work assignment?'' In the case of work evaluation this item was: ``How much do you trust that `the term' makes a good-quality work evaluation?'' Participants responded to this item on the same scale used by \citet{lee2018understanding} with a scale from 1 (do not trust at all) to 7 (extremely trust).

Procedural Justice was measured with seven items taken from \citet{colquitt2001dimensionality} (Cronbach's $\alpha$ = .72). A sample item was ``Have those procedures been free of bias?'' Participants responded to these items on the original scale from 1 (to a very small extent) to 5 (to a very large extent). Note that this scale was not captured in Lee's study.

We again measured affinity for technology by \citet{franke2019personal} as a possible control variable with the same items as in Study 1 (Cronbach's $\alpha$ = .84).~\footnote{Participants were also asked to respond to the question ``In your own words, please briefly explain what you think ‘the term’ is.'', and to report their knowledge or the respective term with one item. Results can be made available upon request.}


\section{Results} \label{sec:results}
\subsection{Study 1 Results} 
\subsubsection{Perceptions regarding the properties of ADM systems}

\begin{figure*}[t]
  \centering
  \includegraphics[width=2\columnwidth]{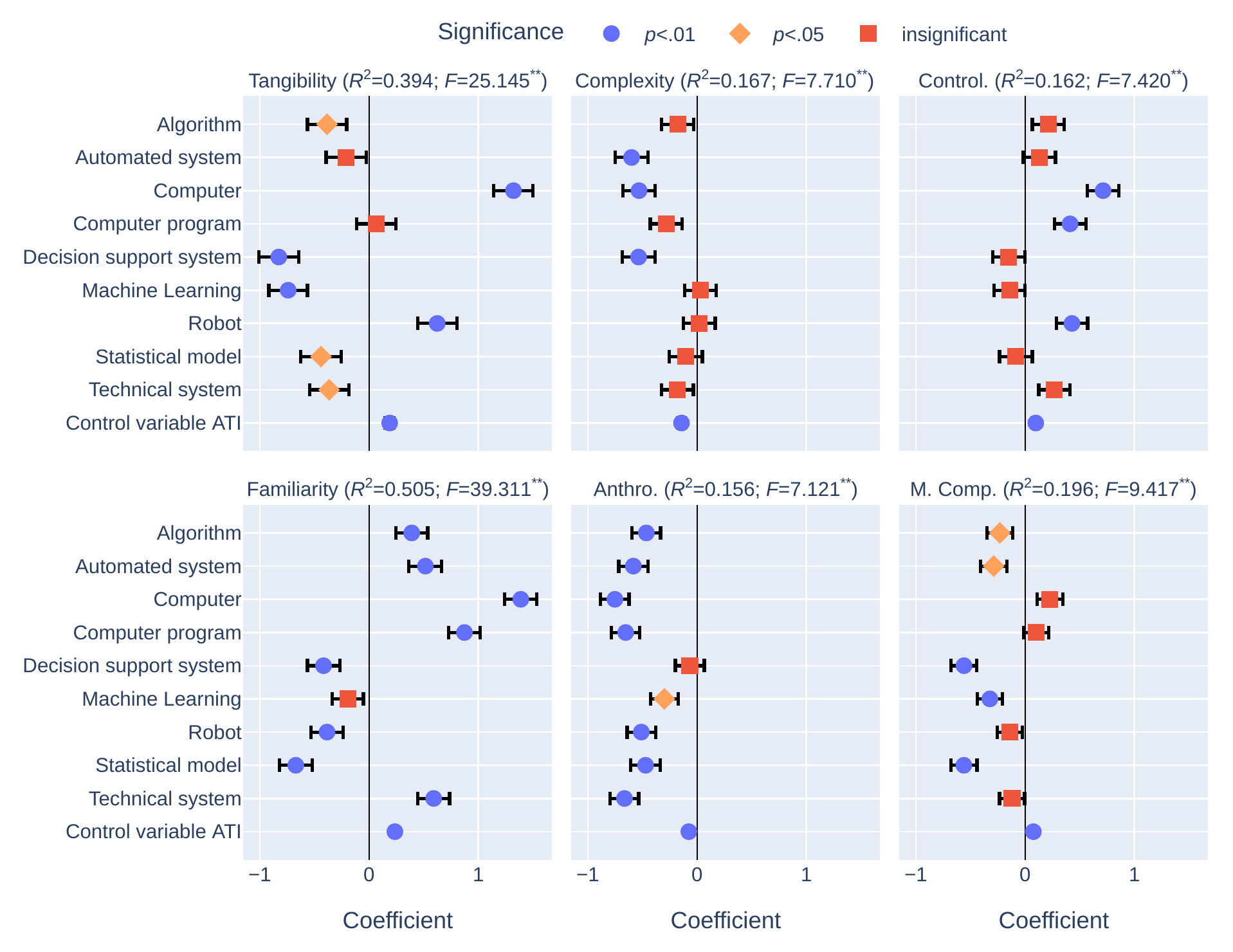}
  \caption{Linear regression coefficient plots for the perceptions of properties of ADM systems depending on the different terms. Dependent variables: Tangibility, Complexity, Controllability (Control.), Familiarity, Anthropomorphism (Anthro.), Machine Competence (M. Comp.). Independent variables: different terms used for ADM systems, and Affinity for technology (ATI) as control variable. The points show the estimated coefficients and respective standard errors. The effects for the terms can be interpreted in comparison to the reference group artificial intelligence (e.g., in the graph for Tangibility, all terms for which the coefficients are displayed on the right side of the black Zero-line received higher ratings for Tangibility than the term artificial intelligence, all left of the line received lower ratings). \textit{R}$^2$ and \textit{F} values were calculated for the respective full model. Mean values and standard deviations for the results can be found in Appendix Table \ref{tab:descriptives_properties}. The intercept of the regression in the figure was omitted for readability purposes, and can be found in Appendix Table \ref{tab:results_regression_study1} that shows the results of the regressions in table format.**\textit{p} < .01. \textit{N} = 397.}
  \label{fig:regression_properties}
  \Description{This figure is fully described in the text and the values for the parameters displayed in this figure can also be found in Appendix Table 6.}
\end{figure*}

Research Question 1 asked whether varying the terminology regarding ADM systems affects perceptions about the properties of ADM systems. To analyze our data, we utilized linear regressions. For each of the six properties (i.e., tangibility, complexity, controllability, familiarity, anthropomorphism, machine complexity), we used separate linear regression models, each including one of the properties as the dependent variable. As independent variables, we used the 10 terms, dummy-coded with the term artificial intelligence as reference group.~\footnote{This was done because in a pilot study, the term artificial intelligence was associated with the highest machine competence and with the comparably highest potential to perform well on the thirteen tasks we included in Study 1.} To be clear, this means we included nine dummy-coded variables into the regression models, where in the first dummy-coded variable the term algorithm was coded with 1, and all other terms were coded with 0, in the second dummy-coded variable the term automated system was coded with 1 and all other terms with 0. In the end, each term is represented in one variable with the coding 1, except for the term artificial intelligence which always received the coding 0 to remain the reference group to which all other groups will be compared. The results regarding how the respective terminology affected the perceived properties of ADM systems are presented in Figure \ref{fig:regression_properties} and can be interpreted in comparison to the reference group artificial intelligence (e.g., how do familiarity perceptions differ between the term artificial intelligence and the term computer program). We additionally entered education level, gender, as well as mean-centered versions of the variables age and affinity for technology as control variables in the regression in order to test whether they affect our results. Education level, age, and gender only showed minor effects on the results which is why we did not include these variables in our final models. However, we included affinity for technology because it was consistently correlated with participants' perceptions of the properties of ADM systems. Participants with a higher affinity for technology perceived ADM systems to be more tangible, less complex, more controllable, found them to be more familiar, were less likely to anthropomorphize systems, and ascribed higher machine competence.~\footnote{Our experimental design allows us to reason about the causal effects of terminology on the dependent variables. However, this is not the case for the control variable affinity for technology. Since this control variable does not reflect an experimental condition that participants were randomly assigned to but instead a characteristic of participants, we can only make claims about the correlation between affinity for technology and the dependent variables.} (This is reflected in Figure \ref{fig:regression_properties}, where the last row of each graph presents the regression weight for affinity for technology; this means if the dot for affinity for technology is right to the zero line, affinity for technology was positively associated with the respective property, if it is left to the zero line, it was negatively associated with the respective property.) The final set of variables in our models thus included the nine dummy-coded variables that reflect the comparison of the respective terms to the reference group artificial intelligence as well as the control variable affinity for technology.

Results for \textbf{tangibility} showed that computers and robots were perceived as more tangible than artificial intelligence. In contrast, decision support systems, machine learning, sophisticated statistical models, algorithms, and technical system were perceived as comparably less tangible. For \textbf{complexity}, results indicated that the term artificial intelligence is perceived to be associated with an entity that is more complex than automated systems, decision support systems, and computers. Regarding \textbf{controllability}, results showed that computers, robots and computer programs were perceived as more controllable than artificial intelligence. Results for \textbf{familiarity} revealed that people perceived computers and computer programs to be especially more familiar than artificial intelligence, but also algorithms, automated systems, and technical systems. In contrast, sophisticated statistical models and robots were perceived as less familiar. For \textbf{anthropomorphism}, our findings showed that the term artificial intelligence was more strongly anthropomorphized than the majority of the other terms, especially computers, computer programs, technical system and automated systems. Finally, participants associated relatively high \textbf{machine competence} with artificial intelligence, computers, and computer programs whereas they perceived especially less machine competence for decision support systems and sophisticated statistical models. Also machine learning, automated systems, and algorithms were perceived as having less machine competence than artificial intelligence.

Another aspect that revealed the influence of terminological differences is the \textit{R$^{2}$} statistic found for the properties associated with ADM systems (see Figure \ref{fig:regression_properties}). This statistic reveals how much variance in participant responses can be explained by the included predictors. In the case of familiarity this means that 50\% of variance is explained by affinity for technology and the terminological differences. When excluding affinity for technology, there was 44\% explained variance due to different terminology. For the other variables, excluding affinity for technology from the model resulted in 36\% explained variance for tangibility, 18\% for machine competence, 14\% for controllability and anthropomorphism, and 11\% for complexity. These results revealed that the strength of the effect of terminological differences varies depending on the respective properties.

In summary, in response to Research Question 1, terminological differences do affect people's perceptions regarding the properties of ADM systems. This seems to be true for all the variables we captured in Study 1. We found the strongest effect of terminology on familiarity and tangibility, and less strong but still significant effects for machine competence, anthropomorphism, controllability and complexity.

\begin{table}[t]
 \centering
 \caption{Results of the linear mixed model with a participant random-effects term, for the comparison of whether humans or systems are better able to conduct a respective task depending on the different tasks and the different terminology.}
 \label{tab:hlm_results}
\begin{tabular}{ll} 
\hline 
\\ [-2.8ex] & Better than human \\ 
  & \textit{Estimates} (\textit{SE}) \\
\hline
 \rule{0pt}{3ex}\hspace{-2mm}  
 Constant & \phantom{$-$}3.671$^{**}$ (0.104) \\ 
  \rule{0pt}{3ex}\hspace{-2mm} 
 \textbf{Within-participant effects} & \\ 
 \rule{0pt}{3ex} 
 Predict weather & \phantom{$-$}0.108\phantom{**} (0.063) \\ 
 \rule{0pt}{3ex} 
 Make work schedules & $-$0.005\phantom{**} (0.063) \\ 
 \rule{0pt}{3ex} 
 Predict the spread of infectious diseases & $-$0.060\phantom{**} (0.063) \\ 
 \rule{0pt}{3ex} 
 Assess dangerous situations while driving & $-$0.428$^{**}$ (0.063) \\ 
 \rule{0pt}{3ex} 
 Evaluate X-rays and MRIs & $-$0.542$^{**}$ (0.063) \\ 
 \rule{0pt}{3ex} 
 Shopping recommendations & $-$0.657$^{**}$ (0.063) \\ 
 \rule{0pt}{3ex} 
 Evaluate applicant documents & $-$1.010$^{**}$ (0.063) \\ 
 \rule{0pt}{3ex} 
 Make medical diagnoses & $-$1.076$^{**}$ (0.063) \\ 
 \rule{0pt}{3ex} 
 Make recidivism predictions & $-$1.121$^{**}$ (0.063) \\ 
 \rule{0pt}{3ex} 
 Therapy recommendations in medicine & $-$1.214$^{**}$ (0.063) \\ 
 \rule{0pt}{3ex} 
 Evaluate job interviews & $-$1.607$^{**}$ (0.063) \\ 
 \rule{0pt}{3ex} 
 Diagnose mental illnesses & $-$1.728$^{**}$ (0.063) \\ 
 \rule{0pt}{3ex}\hspace{-2mm}  
  \textbf{Between-participants effects} & \\ 
 \rule{0pt}{3ex} 
 Algorithm & \phantom{$-$}0.039\phantom{**} (0.135) \\ 
 \rule{0pt}{3ex} 
 Automated system & \phantom{$-$}0.002\phantom{**} (0.139) \\ 
 \rule{0pt}{3ex} 
 Computer & \phantom{$-$}0.044\phantom{**} (0.135) \\ 
 \rule{0pt}{3ex} 
 Computer program & \phantom{$-$}0.011\phantom{**} (0.134) \\ 
 \rule{0pt}{3ex} 
 Decision support system & $-$0.140\phantom{**} (0.137) \\ 
 \rule{0pt}{3ex} 
 Machine learning & \phantom{$-$}0.023\phantom{**} (0.133) \\
 \rule{0pt}{3ex} 
 Robot & $-$0.128\phantom{**} (0.135) \\
 \rule{0pt}{3ex} 
 Statistical model & $-$0.007\phantom{**} (0.140) \\ 
 \rule{0pt}{3ex} 
 Technical system & $-$0.039\phantom{**} (0.134) \\ 
  \rule{0pt}{3ex}\hspace{-2mm}  
  \textbf{Control Variable} & \\ 
   \rule{0pt}{3ex} 
 Affinity for technology & \phantom{$-$}0.136$^{**}$ (0.030) \\
\hline
\multicolumn{2}{l} {\textit{Note:} Higher values for the dependent variable ``Better than }\\
\multicolumn{2}{l} {human'' indicate that participants believed systems to perform  } \\ 
\multicolumn{2}{l} {better than humans. The results for the tasks can be interpreted } \\
\multicolumn{2}{l} {in comparison to the task identify faces. The results for the }\\
\multicolumn{2}{l} {terms can be interpreted in comparison to the term artificial }\\
 \multicolumn{2}{l}{intelligence. The column ``Better than human'' shows estimates }\\
  \multicolumn{2}{l}{and respective  standard errors (SE) in brackets.}\\
\multicolumn{2}{l} {*\textit{p} < .05, **\textit{p} < .01. \textit{N} = 397.}
\end{tabular}
\end{table}

\subsubsection{Evaluations regarding the ability to conduct different tasks}

Research Question 2 asked whether varying the term to refer to ADM systems affects people’s evaluation regarding the performance of systems in various tasks. To investigate this research question, we used a linear mixed model, with a random-effects term for participants. We have included this random-effects term because every participant provided their evaluation of all thirteen tasks, thus the evaluation of tasks was nested within participants. Since all participants evaluated the performance of ADM systems in comparison to humans for all thirteen tasks, we added the tasks as dummy-coded independent variables into the model to investigate within-participant differences in reactions regarding whether humans or systems are better able to conduct a respective task. We used ``identify faces'' as the reference task.~\footnote{We did this because a pilot study indicated that people associate the comparably highest performance with ADM systems that identify faces.} Furthermore, we included the ten terms dummy-coded with the term artificial intelligence as the reference group. We finally also added the mean-centered version of affinity for technology as control variable. The final set of variables in this model thus included the twelve dummy-coded variables for the different tasks, the nine dummy-coded variables for the different terms, and affinity for technology (for which we found that participants with a higher level of affinity for technology evaluated the performance of ADM systems more positively, see \ref{tab:hlm_results}) all to predict the evaluation of the performance of ADM systems in comparison to humans (i.e. the dependent variable ``Better than human'').

\begin{figure}[t]
  \centering
  \includegraphics[width=\columnwidth]{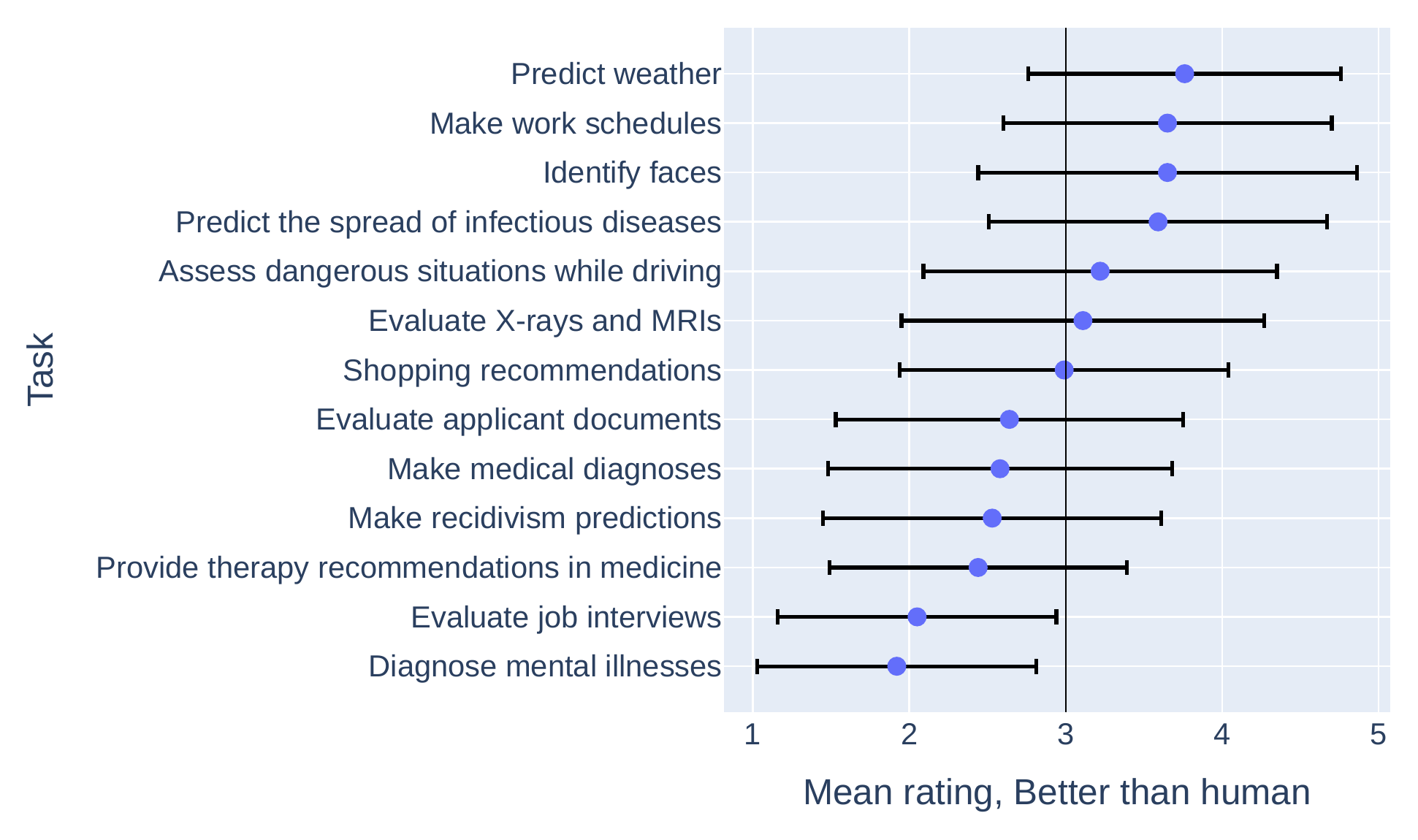}
  \caption{Rank order of participant evaluation of the comparison between humans and systems for the tasks presented in Study 1. A mean of 3 would mean that humans and systems can perform the task equally well (the black line reflects the mean of 3), above 3 means that systems are evaluated to perform better than humans, under 3 means that humans are evaluated as performing better than systems. \textit{N} = 397.}
  \label{fig:mean_tasks}
  \Description{This figure is fully described in the text.}
\end{figure}

Table \ref{tab:hlm_results} displays the results of the linear mixed model. There were no significant differences between the term artificial intelligence and the other terms. Consequently, in response to Research Question 2, varying the term did not significantly affect participant's evaluation of whether humans or the respective ADM system is better able to conduct various tasks. However, we need to highlight that in our case, participants were asked to explicitly compare the potential performance of an ADM system to a human being. This comparison might have reduced the possible effect of terminology because it may have affected how people think about the ADM system in question. With the explicit comparison to humans, people might reduce any term associated with ADM systems to a monolithic concept ``not a human'' instead of using the term to more elaborately think about the system in question. In contrast, it is conceivable that results would have differed if participants reported on how well they believe an, for instance, algorithm would be able to conduct a task on a scale from 1 (not at all) to 5 (very well).

Recent work shows emerging interest in understanding the tasks where people believe systems to perform better or at least equally well as human beings (see e.g., \cite{castelo2019task, dietvorst2020people, langer2021future, lee2018understanding}). Thus, the differences we found between the tasks might be of additional interest to readers. Figure \ref{fig:mean_tasks} presents the mean values and standard deviations for the tasks ranked from most likely to be well-performed by systems to least likely. Our participants were most convinced that systems can perform better than humans for the tasks of predicting weather, identifying faces, scheduling, and predicting the spread of infectious diseases. However, people were convinced that humans perform especially better in the tasks of diagnosing mental illnesses, evaluating job interviews, and providing therapy recommendations in medicine.

\subsection{Study 2 Results} 

\begin{figure}[t]
  \centering
  \includegraphics[width=\columnwidth]{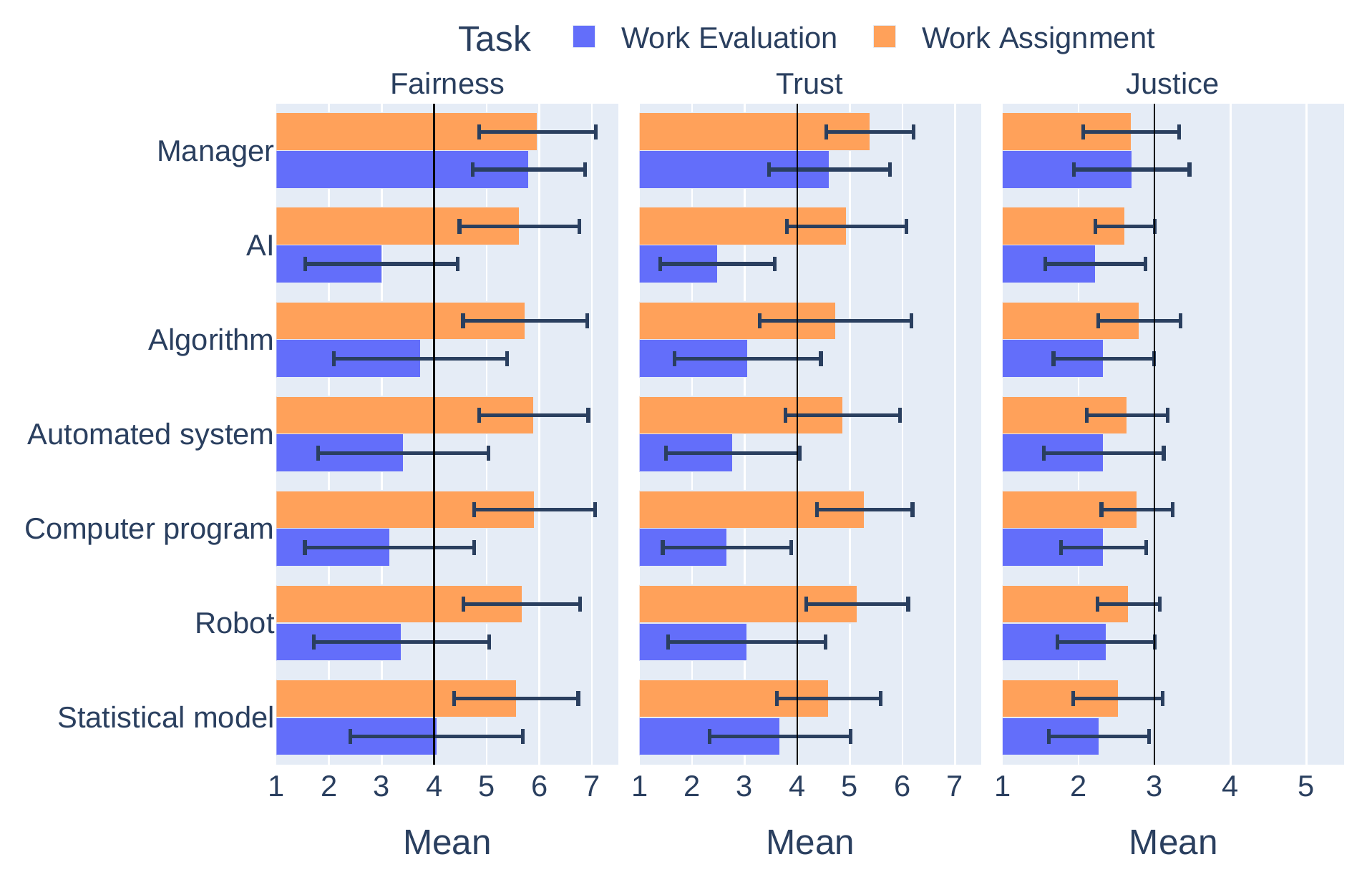}
  \caption{Means and standard deviations for fairness, trust, and justice depending on the work evaluation and work assignment task and depending on the different terminology. The black line reflects the mean point of the respective scale and error bars reflect the standard deviation. The contents of this table can also be found in Appendix Table \ref{tab:descriptives_study2}.}.
  \label{fig:descriptives_study2}
  \Description{The contents of this figure are also displayed in Appendix Table 7.}
\end{figure}

Research Question 3 asked whether varying the term to refer to ADM systems affects people's evaluations of ADM systems. Furthermore, Research Question 4 asked whether the term used to refer to ADM systems and the task for which ADM systems are used interact to affect evaluations of ADM system. Figure \ref{fig:descriptives_study2} provides an overview on means and standard deviations for the dependent variables depending on the tasks and terms.

\begin{figure*}[t]
  \centering
  \includegraphics[width=2\columnwidth]{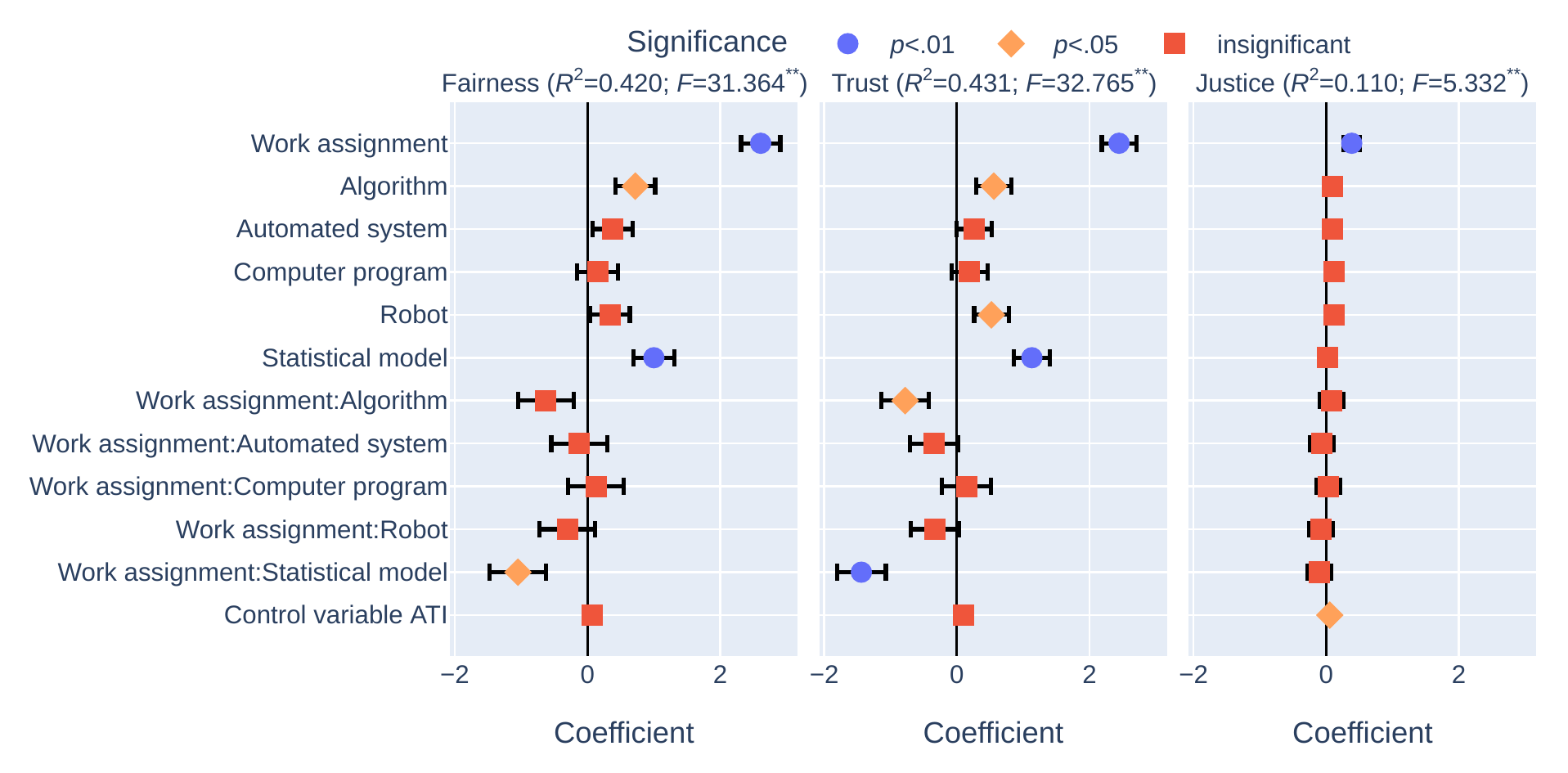}
  \caption{Linear regression coefficient plots for the evaluation of Fairness, Trust, and Procedural Justice depending on the different terms and tasks, including interaction effects. Dependent variables: Fairness, Trust, and Justice. Independent variables: term (six different terms for ADM systems), task (work evaluation or work assignment), and Affinity for technology (ATI) as control variable. The points show the estimated coefficients and respective standard errors. The effects for the terms can be interpreted in comparison to the reference group artificial intelligence, the effect for the tasks can be interpreted in comparison to the reference group work evaluation. \textit{R}$^2$ and \textit{F} values were calculated for the respective full model. The intercept is omitted for readability purposes, and can be found in Appendix Table \ref{tab:regression_study2}. **\textit{p} < .01. \textit{N} = 533.}
  \label{fig:regression_study2}
  \Description{This figure is fully described in the text and the values for the parameters displayed in this figure can also be found in Appendix Table 8.}
\end{figure*}

To investigate Research Questions 3 and 4, we used three separate linear regressions where we included fairness, trust, and procedural justice as dependent variable respectively. For the independent variable ``task'' (i.e., work evaluation vs. work assignment), we entered the tasks as a dummy-coded variable into the regression with work evaluation as the reference group. For the independent variable ``terms'', we included the six different terms using dummy-coded variables in the regression with the term artificial intelligence as reference group. To be able to respond to Research Question 4 that proposed a possible interaction effect between tasks and terms, we added the interaction between the tasks and the terms into the regression model. We finally added the mean-centered version of affinity for technology as control variable. The final set of variables in this model thus included a single dummy-coded variable for the tasks, five dummy-coded variables for the terms, five variables for the interaction between tasks and terms, and the control variable affinity for technology (which was positively associated with perceived justice).

Figure \ref{fig:regression_study2} displays the results for the regressions. Results showed that trust, fairness, and procedural justice were all stronger for ADM systems conducting the task work assignment compared to work evaluation. This replicates Lee' \cite{lee2018understanding} results. Furthermore, the terms algorithm and sophisticated statistical model led to, overall, better fairness evaluations compared to the term artificial intelligence. Additionally, the terms algorithm, robot and sophisticated statistical model led to, overall, higher trust compared to the term artificial intelligence. However, terminological differences did not affect procedural justice evaluations. One interpretation for this finding could be that we used a multiple-item measure for the assessment of justice \cite{colquitt2001dimensionality} whereas for fairness and trust evaluations we used the single-item assessments of these constructs also used in the original study by \citet{lee2018understanding}. For multiple-item measures, the influence of terminological differences might be weaker because when building a scale-mean over multiple items, terminological differences may average out. For instance, maybe the term algorithm in comparison to artificial intelligence leads to a higher evaluation for item one, but a lower evaluation for items two and three. Such a potential effect where the impact of terminology averages out over multiple items is of course not possible for single-item measures. In sum, in response to Research Question 3, varying the term to refer to an ADM system can affect evaluations of ADM systems. In our case, it affected fairness and trust but not procedural justice evaluations. Additionally, it might be that effects of terminological differences depend on the operationalization of the dependent variables.

In response to Research Question 4, Figure \ref{fig:regression_study2} additionally reveals that there were significant interactions for the term sophisticated statistical model and the tasks for fairness and trust. These interaction effects reflect the finding that the term sophisticated statistical model led to the most favorable fairness and trust evaluations for the task work evaluation but to the least favorable fairness and trust evaluations for the task work assignment. Apparently, terminological differences may not only affect fairness and trust evaluations but may also affect such evaluations differently for different tasks. Therefore, in response to Research Question 4, the task and the term may interact to differently affect people's evaluations of ADM systems -- whereas in one task a term may be associated with comparably positive evaluations, this may not hold for another task.

\begin{figure*}
  \centering
  \includegraphics[width=2\columnwidth]{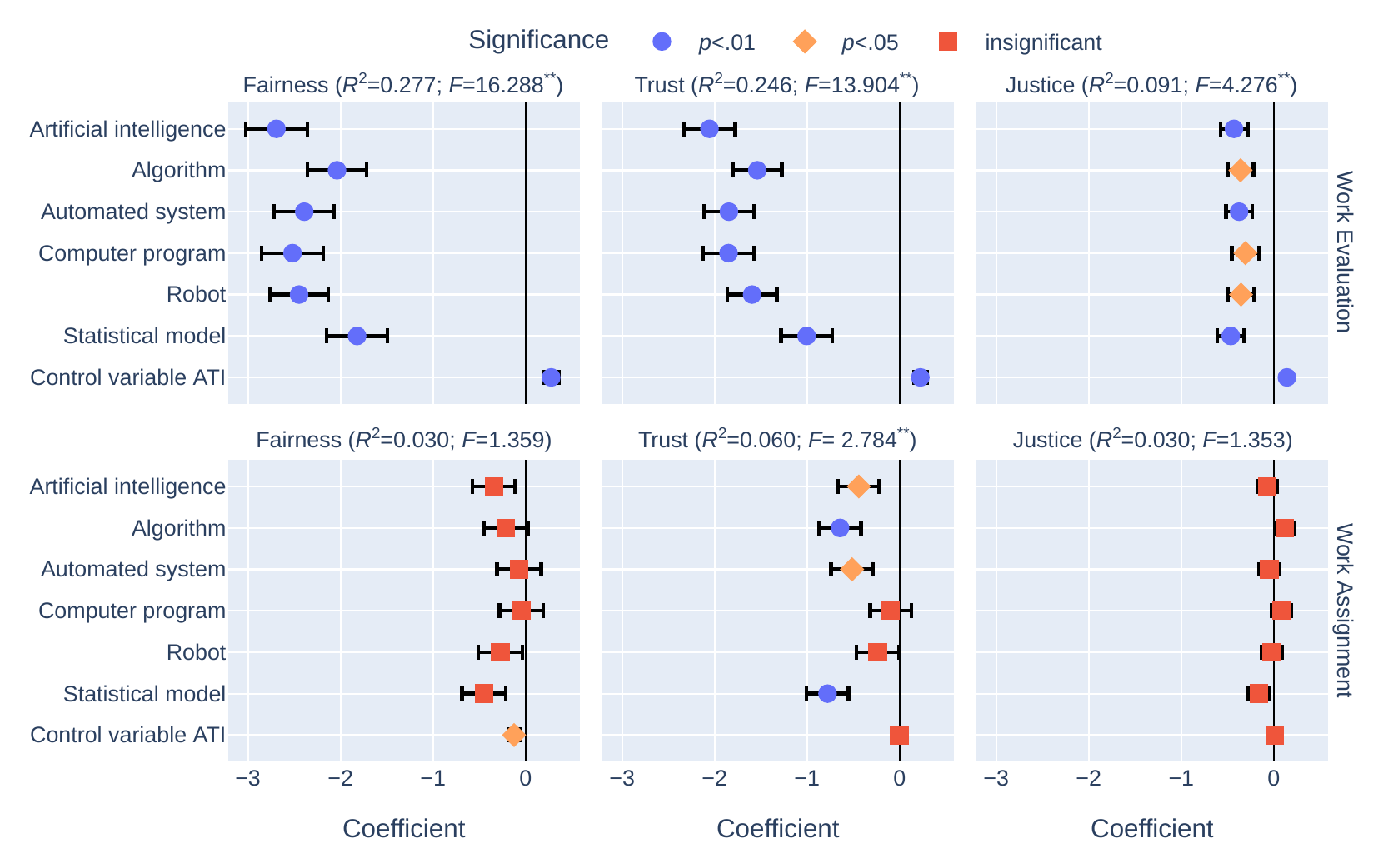}
  \caption{Linear regression coefficient plots for the evaluation of Fairness, Trust, and Procedural Justice for the tasks work evaluation (top) and assignment (bottom) depending on the different terms. Dependent variables: Fairness, Trust, and Justice. Independent variables: human manager vs. different terms for ADM systems, and Affinity for technology (ATI) as control variable. The points show the estimated coefficients and respective standard errors. The effects for the terms can be interpreted in comparison to the human manager as a reference group (e.g., in the graph for Fairness, all terms for which the coefficients are displayed on the left side of the black zero-line received lower ratings for Fairness than the human manager). \textit{R}$^2$ and \textit{F} values were calculated for the respective full model. The intercept is omitted for readability purposes, and can be found in Appendix Tables \ref{tab:regression_study2_evaluation} and \ref{tab:regression_study2_assignment}.\\ **\textit{p} < .01. Work evaluation \textit{n} = 306, Work assignment \textit{n} = 316. }
  \label{fig:regression_work}
  \Description{This figure is fully described in the text and the values for the parameters displayed in this figure can also be found in Appendix Table 9 for the work evaluation task which is the top half of this figure and Table 10 for the work assignment which is the bottom half of this figure.}
\end{figure*}

Finally, in a direct replication of \citet{lee2018understanding}, we investigated the difference between the human manager and ADM systems depending on the tasks as well as on the different terms to refer to ADM systems. Lee found that for work evaluation, her participants evaluated algorithms to be less fair and reported lower trust in algorithms than in human managers. In contrast, the differences she found between human managers and algorithms were much smaller, and non-significant, for work assignment tasks. In order to increase the interpretability of our results in comparison to Lee's results, we followed her example and provide analyses for the tasks separately in Figure \ref{fig:regression_work}. For these linear regressions, we entered six dummy-coded variables with the reference group human manager. We again added affinity for technology as control variable. The final set of variables in our models thus included six dummy-coded variables that reflect the comparison between human manager and the different terms, as well as the control variable affinity for technology (for affinity for technology see last row of each graph in Figure \ref{fig:regression_work}; affinity for technology was positively associated with fairness, trust and justice in the work evaluation task and negatively associated with fairness in the work assignment task). For transparency purposes, we also provide results for procedural justice but will not discuss them as Lee did not measure justice in her study.

Our results for the work evaluation task (Figure \ref{fig:regression_work} top) showed that, although the terms differed in the extent to which they were evaluated as less fair and in the extent to which they evoked less trust (e.g., for the term sophisticated statistical model, there was a smaller difference to the human manager compared to the term artificial intelligence), the differences between human manager and all the terms were significant. This replicates the findings from Lee. 

Yet, results for work assignment (Figure \ref{fig:regression_work} bottom) only partly replicated Lee's findings. Specifically, our results replicated her findings with respect to fairness, where we also found no significant differences between the human manager and the ADM system for any of the terms. However, our results regarding trust replicated Lee's findings only for certain terminology. Specifically, there were no significant differences regarding trust for the terms computer program or robot which replicates Lee's findings. In contrast, for the terms artificial intelligence, algorithm, automated system, and sophisticated statistical model, we found significantly lower trust evaluations compared to the human manager. In other words, if we would have chosen one of the former terms, we would have found no statistically significant results, thus supporting Lee's findings. Instead, if we would have tried to replicate her study with one of the latter terms, we would have found significant differences between humans and ADM systems for trust regarding conducting work assignment tasks and would have concluded that we could not replicate Lee's findings. Consequently, the choice for or against one of the terms could have crucially affected whether our study would have supported or contradicted Lee's results. Similarly, if Lee would have chosen a different term for her study, she might have found different results and might have drawn different conclusions.

\section{Discussion} \label{sec:discussion}

The goal of this paper was to investigate whether terminological differences affect human perceptions and evaluations of ADM systems. The main results are that, indeed, the terminology used to describe ADM systems affects people's perceptions of the properties of those systems. Furthermore, although terminological differences did not affect evaluations of the ability of ADM systems to conduct different tasks in comparison to humans, it did affect people's evaluations of system fairness and trust in systems. These effects might depend on the ways we measure perceptions and evaluations of ADM systems -- in our case we hypothesize that it may depend on whether there was an explicit comparison to human task performance, and on whether we used single-item versus multiple-item measures. However, further research is necessary to evaluate these hypotheses. Overall, we conclude that terminology matters when describing ADM systems to participants in research studies because it can affect the robustness and replicability of research results, and terminology matters because it may shape perceptions and evaluations of ADM systems in communication about such systems (e.g., in public discourse and policy-making). Consequently, it is necessary to be aware that choosing the terms to describe ADM systems can have unintended consequences (e.g., varying research findings due to varying terminology) but that terminology can also be used strategically (e.g., referring to a system as artificial intelligence to make it sound complex and novel).

\subsection{Terminology affects human perceptions and evaluations of ADM systems}

One of the main implications of our study is that it is necessary to be mindful regarding what term to use when describing ADM systems to research participants because findings may vary due to using different terminology. Our Study 2 supports that this might have been an issue in previous HCI research and thus is in line with Langer and Landers' \cite{langer2021future} conclusion that terminological differences may have led to different conclusions for studies that examined similar research questions. For instance, whereas \citet{lee2018understanding} used the term algorithm to describe a system in a hiring scenario, \citet{marcinkowski2020implications} used the term AI technology for a similar task. Whereas Lee and Marcinkowski et al. may have had a similar idea as well as a similar technology in mind -- a system that automatically evaluates applicant information and recommends rejection or approval of applicants -- Lee found that her participants preferred the human manager over the algorithm in hiring, whereas Marcinkowski et al. reported that their participants preferred the AI technology over a human. Part of these differences in findings may be due to the varying terminology (see also \cite{hou2021expert}). Unfortunately, we cannot conclude that there is a simple main effect of different terminology where one term will always lead to more favorable evaluations than another. More precisely, our Study 2 showed that algorithms were perceived as more favorably than artificial intelligence to conduct work evaluations, whereas in the comparison of Lee and Marcinkowski et al.'s results, the term AI technology was associated with more favorable evaluations of ADM systems than the term algorithm. This suggests that terminological differences may differentially affect the evaluation of ADM systems for various tasks. 

Our Study 2 further supports this interpretation because we found that the effect of the terminology depended on the task for which a system is used. Given that systems were perceived more negatively for work evaluation tasks, a preliminary interpretation of this finding might be that we can expect stronger effects of terminology in contexts where people are less positive about the use of systems. This could be the case because in tasks where people already have positive views about systems, they might already expect that ADM systems conduct respective tasks. However, in tasks where it is more controversial whether and to what extent we can and should use ADM systems (e.g., in work evaluation, diagnosis of mental illnesses; \cite{langer2021future}), people's expectations will be violated by the fact that ``not a human'' is conducting the respective task. In cases where expectations are violated, people might be more critical, may think more intensely about the use of systems in those situations, and may consequently scrutinize available information in order to determine how positive or negative they find the idea of an ADM system conducting a task \cite{smith1996message}. In Study 2, this applies to the work evaluation task that was also found to be less positively evaluated by Lee's \cite{lee2018understanding} participants and where our participants may have more intensely thought about what the respective term would tell them about the system conducting this task. In contrast, for work assignment our participants may have been less surprised by the fact that an ADM system conducts the task which led to less elaboration about the term used to describe the system. The mean values and standard deviations found for the single tasks and terms in Study 2 may support this interpretation (see Figure \ref{fig:descriptives_study2} and Appendix Table \ref{tab:descriptives_study2}). Specifically, mean values for fairness in the work evaluation task showed stronger variance and ranged between 3.00 and 4.05, whereas those for the work assignment task ranged between 5.56 and 5.91. The same was found for trust, where the range for work evaluation was between 2.48 and 3.67, whereas for work assignment it was between 4.60 and 5.28. Also, the standard deviations for the evaluation of trust and fairness were almost consistently higher for the work evaluation task. This means that there was more variation between people in how (un)favorably they evaluated the terms for the task work evaluation compared to the task work assignment. Nevertheless, readers should be aware that this is a tentative interpretation of our findings. Shedding further light on the conditions that affect the strength of the influence of different terminology will be a task for future research.

Importantly, we do not claim that terminology is the only factor that contributes to variation in findings and conclusions between studies or that terminology is an especially strong determinant of research findings. In fact, there are many other important choices in studies that will have a larger effect on participants. For instance, both our studies showed that choices regarding the operationalization of constructs (e.g., single-item versus multiple-item measures; explicit comparison to human performance) can influence results. Moreover, both our studies support prior work suggesting that the task for which an ADM system is used more strongly affects participants' evaluations of ADM systems \cite{castelo2019task,dietvorst2020people,langer2021future,lee2018understanding,longoni2019resistance} than terminological differences. For instance, the rank order we found in Study 1 and also the large differences between the tasks in Study 2 support previous work where authors suspected that in high-stakes tasks (e.g., diagnosing mental illnesses), people will find humans to be better suited to perform these tasks than ADM systems \cite{langer2021future, langer2019highly}. Yet, high-stakes versus low-stakes is clearly not the only dimension that explains differences in evaluations of ADM systems in these tasks. For example, predicting the spread of infectious diseases might also be considered a high-stakes task (especially given that our data collection was conducted during the COVID-19 pandemic) and our participants evaluated that ADM systems would be better able to perform this task than humans. Other dimensions might be whether the task requires human versus mechanical skills \cite{lee2018understanding}, the inherent uncertainty associated with the decision-making task at hand \cite{dietvorst2020people}, and the complexity of the task \cite{nagtegaal2021impact} (as has been argued by \citet{langer2021future}). 

Moreover, providing specific information on characteristics of ADM systems may reduce effects that terminology may have. Specifically, terminology effects might stem from what participants have in mind when thinking about an ADM system described with a specific terminology. As we described in Section \ref{sec:terminology_matters}, a large share of previous work only used the respective term to inform participants that an ADM system will make or support decisions without further information regarding different nuances of underlying technology or regarding how well a system works for a specific task. With this kind of ambiguity, terminology effects may be especially strong. However, providing information on, for example, training and validation of an ADM system or describing that during validation the respective system has been found to make accurate predictions in 95\% of cases may attenuate terminology effects because participants are not left wondering how a system was developed or how well a system will work. In cases where there are more specific descriptions of system characteristics, it will be a task for future research to investigate whether it matters less that a system is described with different terminology.

\subsection{Being mindful about terminology may enhance robustness and replicability of research}

Even if the effects of terminology may depend on other methodological choices (i.e., the choice of operationalization of constructs), are comparably weaker than the effects of other considerations (e.g., the task performed by ADM systems), or are attenuated under certain circumstances (e.g., when adding specific information about characteristics of ADM systems) our results showed that different terminology is associated with variation in people's perceptions and evaluations of ADM systems. Given that the terminology to describe ADM systems to participants is easily controllable within studies, we suggest that researchers
\begin{itemize}
  \item mindfully consider what term to use to describe ADM systems to their participants
  \item clearly report in the methodology of their papers what term they used
\end{itemize}

Following these suggestions may help increase the robustness and replicability of research findings. More precisely, when designing a study where participants are informed that an ADM system decides about the future of people or in studies where people interact with an ADM system to perform a task, it makes sense to screen previous literature to examine what terms other authors have used to describe the respective systems. If the goal of the research is to replicate or advance specific previous studies, it makes sense to use similar terms like the respective studies since it would at least control for unintended variation due to different terminology. Unfortunately, Langer and Landers' \cite{langer2021future} literature review showed that there is a large variety of terms that have been used in previous studies. To date, there is limited information regarding how strongly varying terminology has affected findings and conclusions of previous work. We hope that our studies raise awareness of the effects different terminology can have on research findings and hope it will motivate future research to more actively consider what term to use when describing ADM systems to participants.

\subsection{Terminology can be used strategically in communication about ADM systems}

Our studies imply that when people communicate with someone about ADM systems describing this system with different terminology can impact listeners' perceptions and evaluations associated with the respective system (for similar results across disciplines see \cite{eitzel2017citizen, puhl2020words, shank2020software}). This supports that terminology may lead to different reactions in communication about ADM systems. An implication that needs further exploration is whether different perceptions and evaluations of ADM systems lead to different behavior in the interaction with ADM systems. For instance, if people are more likely to trust a statistical model compared to an artificial intelligence to conduct work assignment, they may also be more likely to actually use and rely on a system described as being a sophisticated statistical model. Similarly, if people associate higher machine competence with artificial intelligence compared to automated systems, they may more likely use outputs generated by a system that is described as an artificial intelligence. It is important to highlight that these are hypotheses we derived from our studies because we did not measure behavioral outcomes associated with using different terminology. Nevertheless, Study 2 showed that people's fairness and trust evaluations depended on the term used to describe the system and fairness as well as trust have been found to be antecedents of actual system use \cite{hoff2015trust, howard2020implementation, lee2004trust}.

Overall, our studies also suggest that for communication about ADM systems in journalist reports, public discourse, and policy-making it is necessary to be aware that the choice of a certain term has effects. Terminology may affect how people receive the respective communication, how they evaluate the use of ADM systems for various tasks, and may influence what people do as consequence of respective communications. For example, if journalists write about artificial intelligence \cite{windley2021} versus algorithms in recruitment \cite{oren2016howto}, this might lead to different evaluations of the general idea of ADM systems to support recruitment. If the term artificial intelligence would lead to a less favorable evaluation of using ADM systems to evaluate people and make decisions over people's careers (as found in previous work; \cite{langer2021future, lee2018understanding, newman2020eliminating}), this can lead to stronger public outcry and potentially even protests than the idea of algorithms doing the same thing. As another example, if public discourse on healthcare supported by ADM systems would use the term automated system instead of the term artificial intelligence to start respective discussions, this may lead to less engagement and controversy in the discussion because people may already know that automated systems are used in healthcare, thus less likely violating people's expectations.

In conclusion, our study supports that in communication practice, the choice for or against a term can be a strategic one. Take the example of policy-making documents where the authors may have the choice to use the term artificial intelligence compared to more familiar terms such as computer programs. Maybe if the European Commission's ``Ethics Guidelines for Trustworthy AI'' \cite{euai2021} had been called ``Ethics Guidelines for Trustworthy Computer Programs'' there would have been less public outreach. In other words, terminology could be used intentionally to engage people to contribute to discussions. Furthermore, terminology can also be used as a selling argument for companies who use ADM systems. There are reports that many European companies claim to use artificial intelligence in their products but actually never did so \cite{vincent2019}. Our results showed that it might be the complexity as well as strong potential that people associate with artificial intelligence that may underlie this choice of terminology compared to equally plausible terminology. Consequently, in comparison to sometimes unintended consequences of using different terminology (e.g., variation in research findings), terminology can clearly also been used strategically in order to cause desired effects (e.g., engagement, interest; \cite{eitzel2017citizen, puhl2013motivating}). 

\subsection{Limitations}

There are four main limitations to our work. First, all captured data relied on self-reported information from participants. Although this led us to conclude that terminological differences affect perceptions and evaluations of ADM systems, we can only draw tentative conclusions with respect to behavioral consequences -- consequences that have been found in other fields investigating terminological differences \cite{eitzel2017citizen}. For instance, given that the term artificial intelligence was associated with comparably high machine competence, we imagine future studies where participants interact with a real system providing them with recommendations and where researchers capture to what extent participants rely on the recommendations by the system depending on whether the system is described as artificial intelligence or as a sophisticated statistical model \cite{dietvorst2015algorithm}. 
Second, we gathered non-representative samples on Prolific so we cannot generalize our findings to broader populations. However, since many studies in HCI have been and still are conducted via crowdsourcing platforms such as Prolific or MTurk (e.g., \cite{binns2018s, wang2020factors}), we are optimistic that our interpretation that terminology may have affected HCI research findings holds. 
Third, although we included a measure of participants' affinity for technology, participants' reactions to the terms may have been affected by more specific AI or computer science-related experience. However, by randomly assigning participants to the experimental groups, we would expect that there was no group where there was a significantly larger number of participants with a computer science background. Therefore, we believe that the interpretations drawn from our study remain valid. Nevertheless, future research could examine whether people with a strong computer scientific background would react less strongly to different terminology because they know that some of these terms can be used interchangeably, or whether they react more strongly because they are more aware of the nuances that distinguish the terms (e.g., with respect to underlying technology).
Fourth, our studies were conducted in English so results may have been different if we had conducted our studies in other languages. This limitation may inspire future work that investigates whether terminology has different effects in languages other than English. For example, for scholars who conduct their research in the native languages of their participants it might be interesting to investigate whether terminology has effects similar to what we found in our studies. We can imagine that in some languages the respective terminologies (e.g., robots and artificial intelligence) might be more closely related or might be perceived as more different than in the English language. Furthermore, in some languages respective terms may be more common compared to other languages where terms may reflect something more specific, potentially resulting in differences regarding perceptions of familiarity or complexity. Such nuances may lead to different effects of terminology in different languages.

\subsection{Conclusion}

When communicating, there are many terms that can be used to express similar ideas. This also applies to the terms to refer to ADM systems. Different terminology can strongly affect people's thoughts, feelings, and behavior \cite{eitzel2017citizen, puhl2020words}. From a research point of view, our studies showed that it is necessary to be mindful when describing ADM system, especially when trying to better understand how people perceive and evaluate ADM systems, when investigating what people expect from such systems in application contexts, and when examining how people interact with systems in everyday life. From a practical point of view, our studies imply that in communicating about ADM systems in public discourse, the media, and policy-making, there might be strategic choices for different terminology because terminology may have the potential to engage people, make them more interested in a topic, or may lead to positive/negative evaluations of the use of ADM systems. In summary, our studies show that terminology needs to be chosen wisely as it can affect what kind of properties people ascribe to ADM systems, can influence people's perceptions of systems in application contexts, and can affect the robustness and replicability of research findings.

\begin{acks}

Work on this paper and on the studies included in this paper was funded by the Volkswagen Foundation grant Az. 98513 "Explainable Intelligent Systems" (EIS), by the Deutsche Forschungsgemeinschaft (DFG) grant 389792660 as part of TRR248, and by the Bundesministerium für Bildung und Forschung (BMBF) Project "Ophthalmo-AI" grant 16SV8640.

\end{acks}

\bibliographystyle{ACM-Reference-Format}
\bibliography{references}

\pagebreak
\appendix
\onecolumn
\section{Appendix} \label{sec:appendix}

\begin{longtable}{p{2cm}p{10cm}p{4,5cm}}
\caption{Items for Study 1}
    \label{tab:items_study1} \\
    \hline
Scale & Item text & Response format \\
    \hline
    \endhead
Tangibility 
    & When I think of ``the term'', I have a clear picture in mind. & 1 (strongly disagree) to \\
    & When I think of ``the term'', it has a shape. & 5 (strongly agree) \\
Familiarity 
    & ``The term'' is something I encounter in everyday life. & 1 (strongly disagree) to \\
    & ``The term'' is familiar to me. & 5 (strongly agree) \\
    & ``The term'' is something novel. (r) & ~ \\
Complexity 
    & ``The term'' and how it works is easy to understand. (r) & 1 (strongly disagree) to \\
    & ``The term'' is understandable even for laypeople. (r) & 5 (strongly agree) \\
    & ``The term'' is complex. & ~ \\
    &  I could predict the results generated by ``the term''. (r) (e) & ~ \\
    & ``The term'' works, even if I do not exactly understand how. (e) & ~ \\
Controllability 
    & ``The term'' is controllable by humans. & 1 (strongly disagree) to \\
    & ``The term'' and related processes can be controlled. & 5 (strongly agree) \\
    & ``The term'' acts independently. (r) (e) & ~ \\
Anthropo-
    & ``The term'' has a mind on its own. & 1 (strongly disagree) to \\
morphism & ``The term'' has intentions. & 5 (strongly agree) \\
    & ``The term'' has free will. & ~ \\
    & ``The term'' has beliefs. & ~ \\
    & ``The term'' has the ability to experience emotions. & ~ \\
    & ``The term'' has desires. & ~ \\
    & ``The term'' has conscientiousness. & ~ \\
    & ``The term'''s decision making processes are similar to those of humans. & ~ \\
Machine
    & ``The term'' has great potential in terms of what it can be used for. & 1 (strongly disagree) to \\
competence & ``The term'' can be used flexibly for various tasks. & 5 (strongly agree) \\
    & I believe that ``the term'' has great capabilities. & ~ \\
    & ``The term'' can generate results just as good as human experts. & ~ \\
    & ``The term'' can adapt to changing situations. & ~ \\
    & ``The term'' and their decision outcomes are similar to that of humans. & ~ \\
Tasks
    & ``The term'' can make shopping recommendations. & 1 (worse than a human),  \\
    & ``The term'' can evaluate documents by applicants (e.g., applicant resumes).& 2 (slightly worse than a human),  \\
    & ``The term'' can make recidivism predictions of convicted offenders.& 3 (as good as a human), \\
    & ``The term'' can make medical diagnoses.& 4 (slightly better than a human), \\
    & ``The term'' can evaluate X-ray and MRI images.& 5 (better than a human) \\
    & ``The term'' can predict the weather.& ~ \\
    & ``The term'' can evaluate job interviews.& ~ \\
    & ``The term'' can produce shift schedules at work.& ~ \\
    & ``The term'' can provide therapy recommendations in medicine.& ~ \\
    & ``The term'' can diagnose mental illness.& ~ \\
    & ``The term'' can identify faces.& ~ \\
    & ``The term'' can assess dangerous situations while driving.& ~ \\
    & ``The term'' can predict the spread of infectious diseases.& ~ \\
Affinity for
    & I like to occupy myself in greater detail with technical systems. & 1 (completely disagree) to  \\
technology & I like testing the functions of new technical systems. & 6 (completely agree) \\
    & It is enough for me that a technical system works; I don't care how or why. (r) & ~ \\
    & It is enough for me to know the basic functions of a technical system. (r) & ~ \\
    \hline
\multicolumn{3}{l} {\textit{Note:} ``The term'' is used as a placeholder for the experimentally manipulated terms respectively. (r) = reverse-coded item, } \\
\multicolumn{3}{l} {(e) = item was excluded from the final analysis because it led to a low Cronbach's $\alpha$ of the scale.} \\
\end{longtable}

\newpage

\begin{longtable}{p{1,5cm}p{10,5cm}p{4cm}}
\caption{Items for Study 2}
\label{tab:items_study2} \\
\hline
    Scale & Item text & Response format \\
\hline
\endhead
Fairness 
    & How fair or unfair is it for Chris that the ``the term'' assigns him to check a specific component of the machinery and he does the maintenance work on it? / How fair or unfair is it for Chris that the ``the term'' evaluates his performance? & 1 (Very unfair) to 7 (Very fair) \\
Trust
    & How much do you trust that ``the term'' makes a good-quality work assignment? / How much do you trust that ``the term'' makes a good-quality work evaluation? &
    1 (No trust at all) to 7 (Extreme trust) \\
Procedural Justice 
    & The following items refer to the procedures used to arrive at the decision. To what extent do you think: & ~ \\
    & Has Chris been able to express his views and feelings during those procedures? & 1 (to a very small extent), \\
    & Has Chris had influence over the decision arrived at by those procedures? & 2 (to a small extent),  \\
    & Have those procedures been applied consistently? & 3 (to some extent), \\
    & Have those procedures been free of bias? & 4 (to a large extent), \\
    & Have those procedures been based on accurate information? & 5 (to a very large extent) \\
    & Has Chris been able to appeal the decision arrived at by those procedures? & ~ \\
    & Have those procedures upheld ethical and moral standards? & ~ \\
Affinity for
    & I like to occupy myself in greater detail with technical systems. & 1 (completely disagree) to \\
technology & I like testing the functions of new technical systems. & 6 (completely agree) \\
    & It is enough for me that a technical system works; I don't care how or why. (r) & ~ \\
    & It is enough for me to know the basic functions of a technical system. (r) & ~ \\
\hline
\multicolumn{3}{l} {\textit{Note:} ``The term'' is used as a placeholder for the experimentally manipulated terms respectively. (r) = reverse-coded item.} \\
\end{longtable}

\begin{table}[h]
 \centering
 \caption{Means and standard deviations for perceptions regarding the properties of ADM systems for the terms in Study 1.}
 \label{tab:descriptives_properties}
    \begin{tabular}{lcccccccccccc}
  \hline
 Condition & \multicolumn{2}{c}{Tang.} & \multicolumn{2}{c}{Comp.} & \multicolumn{2}{c}{Cont.} & \multicolumn{2}{c}{Fam.} & \multicolumn{2}{c}{Anth.} & \multicolumn{2}{c}{M. Com.} \\
 & \textit{M} & \textit{SD} & \textit{M} & \textit{SD} & \textit{M} & \textit{SD} & \textit{M} & \textit{SD} & \textit{M} & \textit{SD} & \textit{M} & \textit{SD} \\
  \hline
  \rule{0pt}{3ex}
Artificial intelligence & 2.89 & 0.83 & 3.76 & 0.61 & 3.52 & 0.69 & 2.98 & 0.63 & 2.39 & 2.39 & 4.01 & 0.49 \\ 
  \rule{0pt}{3ex} 
Algorithm & 2.41 & 0.63 & 3.66 & 0.65 & 3.69 & 0.78 & 3.26 & 0.68 & 1.97 & 1.97 & 3.74 & 0.54 \\
  \rule{0pt}{3ex} 
Automated system & 2.65 & 0.88 & 3.19 & 0.77 & 3.64 & 0.69 & 3.46 & 0.66 & 1.82 & 1.82 & 3.71 & 0.74 \\
  \rule{0pt}{3ex} 
Computer & 4.19 & 0.78 & 3.25 & 0.87 & 4.22 & 0.59 & 4.34 & 0.62 & 1.65 & 1.65 & 4.23 & 0.41 \\ 
  \rule{0pt}{3ex} 
Computer program & 2.91 & 0.79 & 3.51 & 0.77 & 3.91 & 0.60 & 3.80 & 0.77 & 1.76 & 1.76 & 4.10 & 0.53 \\ 
  \rule{0pt}{3ex} 
DSS & 2.05 & 0.85 & 3.24 & 0.59 & 3.37 & 0.63 & 2.55 & 0.69 & 2.33 & 2.33 & 3.45 & 0.43 \\ 
  \rule{0pt}{3ex} 
Machine learning & 2.15 & 0.83 & 3.79 & 0.59 & 3.38 & 0.73 & 2.79 & 0.69 & 2.09 & 2.09 & 3.69 & 0.54 \\ 
  \rule{0pt}{3ex} 
Robot & 3.46 & 0.82 & 3.82 & 0.67 & 3.93 & 0.69 & 2.54 & 0.74 & 1.90 & 1.90 & 3.85 & 0.46 \\ 
  \rule{0pt}{3ex} 
Statistical model & 2.43 & 0.91 & 3.68 & 0.66 & 3.43 & 0.44 & 2.29 & 0.59 & 1.93 & 1.93 & 3.44 & 0.37 \\
  \rule{0pt}{3ex} 
Technical system & 2.50 & 0.97 & 3.60 & 0.53 & 3.78 & 0.65 & 3.54 & 0.86 & 1.74 & 1.74 & 3.88 & 0.64 \\ 
   \hline
   \multicolumn{13}{l}{\textit{Note:} The columns Tang., Comp., Cont., Fam., Anth., and M. Comp show the mean values for these }\\
   \multicolumn{13}{l}{variables. Tang. = Tangibility, Comp. = Complexity, Cont. = Controllability, Fam. = Familiarity, }\\
   \multicolumn{13}{l}{Anth. = Anthropomorphism, M. Com. = Machine Competence, DSS = Decision support system.}\\
   \textit{N} = 397.
\end{tabular}
\end{table}

\begin{table}[h]
 \centering
 \caption{Results for the linear regressions analyzing the differences between the respective terms for the properties associated with ADM systems.}
 \label{tab:results_regression_study1}
  \setlength\tabcolsep{1.5pt}
\begin{tabular}{@{\extracolsep{5pt}}lcccccc}
\hline
\\[-2.8ex] & Tangibility & Complexity & Controllability & Familiarity & Anthro. & M. Comp. \\ 
\hline
 Constant & 2.861$^{**}$ & 3.786$^{**}$ & 3.510$^{**}$ & 2.947$^{**}$ & 2.405$^{**}$ & 4.001$^{**}$ \\ 
 & (0.127) & (0.103) & (0.102) & (0.103) & (0.092) & (0.081) \\
 
 \rule{0pt}{3ex} 
 Algorithm & $-$0.385$^{*}$ & $-$0.178 & 0.212 & 0.390$^{**}$ & $-$0.465$^{**}$ & $-$0.232$^{*}$ \\ 
 & (0.180) & (0.148) & (0.146) & (0.146) & (0.130) & (0.116) \\ 
 \rule{0pt}{3ex} 
 Automated system & $-$0.209 & $-$0.600$^{**}$ & 0.129 & 0.515$^{**}$ & $-$0.584$^{**}$ & $-$0.288$^{*}$ \\ 
 & (0.185) & (0.151) & (0.149) & (0.150) & (0.134) & (0.119) \\ 
 \rule{0pt}{3ex} 
 Computer & 1.321$^{**}$ & $-$0.532$^{**}$ & 0.713$^{**}$ & 1.388$^{**}$ & $-$0.753$^{**}$ & 0.227 \\ 
 & (0.180) & (0.147) & (0.145) & (0.146) & (0.130) & (0.116) \\ 
 \rule{0pt}{3ex} 
 Computer program & 0.066 & $-$-0.283 & 0.411$^{**}$ & 0.873$^{**}$ & $-$0.654$^{**}$ & 0.102 \\ 
 & (0.179) & (0.146) & (0.144) & (0.145) & (0.130) & (0.115) \\ 
 \rule{0pt}{3ex} 
 Decision support system & $-$0.827$^{**}$ & $-$0.536$^{**}$ & $-$0.150 & $-$0.417$^{**}$ & $-$0.067 & $-$0.560$^{**}$ \\ 
 & (0.182) & (0.149) & (0.147) & (0.148) & (0.132) & (0.117) \\ 
 \rule{0pt}{3ex} 
 Machine Learning & $-$0.741$^{**}$ & 0.028 & $-$0.142 & $-$0.195 & $-$0.300$^{*}$ & $-$0.323$^{**}$ \\ 
 & (0.177) & (0.144) & (0.142) & (0.143) & (0.128) & (0.114) \\ 
 \rule{0pt}{3ex} 
 Robot & 0.623$^{**}$ & 0.019 & 0.428$^{**}$ & $-$0.385$^{**}$ & $-$0.511$^{**}$ & $-$0.139 \\ 
 & (0.179) & (0.146) & (0.144) & (0.145) & (0.130) & (0.115) \\ 
 \rule{0pt}{3ex} 
 Statistical model & $-$0.440$^{*}$ & $-$0.104 & $-$0.085 & $-$0.671$^{**}$ & $-$0.472$^{**}$ & $-$0.561$^{**}$ \\ 
 & (0.186) & (0.152) & (0.150) & (0.151) & (0.135) & (0.120) \\ 
 \rule{0pt}{3ex} 
 Technical system & $-$0.366$^{*}$ & $-$0.181 & 0.268 & 0.591$^{**}$ & $-$0.665$^{**}$ & $-$0.120 \\ 
 & (0.179) & (0.146) & (0.144) & (0.145) & (0.130) & (0.115) \\ 
 \rule{0pt}{3ex}\hspace{-2mm}  
  \textbf{Control Variable} & \\ 
  
   \rule{0pt}{3ex} 
 Affinity for technology & 0.188$^{**}$ & $-$0.143$^{**}$ & 0.097$^{**}$ & 0.236$^{**}$ & $-$0.076$^{**}$ & 0.075$^{**}$ \\ 
 & (0.040) & (0.033) & (0.032) & (0.032) & (0.029) & (0.026) \\ 
\hline
\textit{R$^{2}$} & 0.394 & 0.147 & 0.162 & 0.505 & 0.156 & 0.196 \\ 
\textit{F} & 25.145$^{**}$ & 6.652$^{**}$ & 7.420$^{***}$ & 39.311$^{**}$ & 7.121$^{**}$ & 9.417$^{**}$ \\
& (\textit{df} = 10; 386) & (\textit{df} = 10; 385) & (\textit{df} = 10; 385) & (\textit{df} = 10; 386) & (\textit{df} = 10; 386) & (\textit{df} = 10; 386) \\
\hline 
\multicolumn{7}{l}{\textit{Note:} The effects for the terms can be interpreted in comparison to the reference group artificial intelligence.}\\
\multicolumn{7}{l}{Anthro. = Anthropomorphism, M. Comp. = Machine Competence. The columns Tangibility, Complexity, Controllability,}\\
\multicolumn{7}{l}{Familiarity, Anthro., and M. Comp. show estimates and respective standard errors in brackets.}\\
  \multicolumn{7}{l}{*\textit{p} < .05, **\textit{p} < .01. \textit{N} = 397.} \\
\end{tabular} 
\end{table} 

\begin{table}[h]
 \centering
 \caption{Means and standard deviations for fairness, trust, and justice depending on the work evaluation and work assignment task and depending on the different terminology in Study 2.}
 \label{tab:descriptives_study2}
\begin{tabular}{lccccccc}
 \hline
Condition & \textit{n} & \multicolumn{2}{c}{Fairness} & \multicolumn{2}{c}{Trust} & \multicolumn{2}{c}{Justice} \\ 
 & & \textit{M} & \textit{SD} & \textit{M} & \textit{SD} & \textit{M} & \textit{SD} \\
 \hline 
 \rule{0pt}{3ex} 
Manager, evaluation & 44 & 5.80 & 1.07 & 4.61 & 1.15 & 2.70 & 0.76 \\ 
 \rule{0pt}{3ex} 
Manager, assignment & 45 & 5.96 & 1.11 & 5.38 & 0.83 & 2.69 & 0.63 \\ 
 \rule{0pt}{3ex} 
AI, evaluation & 40 & 3.00 & 1.45 & 2.48 & 1.09 & 2.22 & 0.66 \\ 
 \rule{0pt}{3ex} 
AI, assigmnent & 48 & 5.62 & 1.14 & 4.94 & 1.14 & 2.61 & 0.39 \\ 
 \rule{0pt}{3ex} 
Algorithm, evaluation & 47 & 3.74 & 1.65 & 3.06 & 1.39 & 2.33 & 0.66 \\ 
 \rule{0pt}{3ex} 
Algorithm, assignment & 45 & 5.73 & 1.18 & 4.73 & 1.44 & 2.80 & 0.54 \\ 
 \rule{0pt}{3ex} 
Automated system, evaluation & 44 & 3.41 & 1.62 & 2.77 & 1.27 & 2.33 & 0.79 \\
 \rule{0pt}{3ex} 
Automated system, assignment & 44 & 5.89 & 1.04 & 4.86 & 1.09 & 2.64 & 0.53 \\ 
 \rule{0pt}{3ex} 
Computer program, evaluation & 41 & 3.15 & 1.61 & 2.66 & 1.22 & 2.33 & 0.56 \\ 
 \rule{0pt}{3ex} 
Computer program, assignment & 46 & 5.91 & 1.15 & 5.28 & 0.91 & 2.77 & 0.47 \\ 
 \rule{0pt}{3ex} 
Robot, evaluation & 48 & 3.38 & 1.67 & 3.04 & 1.50 & 2.36 & 0.64 \\
 \rule{0pt}{3ex} 
Robot, assignment & 43 & 5.67 & 1.11 & 5.14 & 0.97 & 2.66 & 0.41 \\ 
 \rule{0pt}{3ex} 
Statistical model, evaluation & 42 & 4.05 & 1.64 & 3.67 & 1.34 & 2.27 & 0.66 \\
 \rule{0pt}{3ex} 
Statistical model, assignment & 45 & 5.56 & 1.18 & 4.60 & 0.99 & 2.52 & 0.59 \\ 
 \hline
  \multicolumn{8}{l}{\textit{Note:} The columns Fairness, Trust, and Justice show the mean values for}\\
    \multicolumn{8}{l}{these variables.}\\
 \multicolumn{8}{l}{\textit{N} = 622.}
 \end{tabular}
\end{table}

\begin{table}[!h] 
 \centering
 \caption{Results of the linear regressions for fairness, trust, and justice evaluations depending on the tasks and the terms in Study 2.}
 \label{tab:regression_study2}
\begin{tabular}{lccc}
\hline 
\\[-2.8ex] & Fairness & Trust & Justice \\ 
\hline
 \rule{0pt}{3ex} \hspace{-2mm}
 Constant & 3.019$^{**}$ & 2.500$^{**}$ & 2.232$^{**}$ \\ 
 & (0.220) & (0.192) & (0.092) \\ 
 \rule{0pt}{3ex} 
 Work assignment & 2.611$^{**}$ & 2.445$^{**}$ & 0.385$^{**}$ \\ 
 & (0.297) & (0.259) & (0.125) \\ 
 \rule{0pt}{3ex} 
 Algorithm & 0.721$^{*}$ & 0.558$^{*}$ & 0.099 \\ 
 & (0.299) & (0.261) & (0.126) \\ 
 \rule{0pt}{3ex} 
 Automated system & 0.380 & 0.260 & 0.089 \\ 
 & (0.304) & (0.265) & (0.128) \\ 
 \rule{0pt}{3ex} 
 Computer program & 0.154 & 0.193 & 0.115 \\ 
 & (0.308) & (0.269) & (0.130) \\ 
 \rule{0pt}{3ex} 
 Robot & 0.340 & 0.521$^{*}$ & 0.117 \\ 
 & (0.298) & (0.260) & (0.125) \\ 
 \rule{0pt}{3ex} 
 Statistical model & 1.001$^{**}$ & 1.131$^{**}$ & 0.020 \\ 
 & (0.309) & (0.269) & (0.130) \\ 
 \rule{0pt}{3ex} 
 Work assignment:Algorithm & $-$0.626 & $-$0.779$^{*}$ & 0.081 \\ 
 & (0.415) & (0.361) & (0.174) \\ 
 \rule{0pt}{3ex} 
 Work assignment:Automated system & $-$0.125 & $-$0.342 & $-$0.070 \\ 
 & (0.420) & (0.366) & (0.176) \\ 
 \rule{0pt}{3ex} 
 Work assignment:Computer program & 0.129 & 0.146 & 0.035 \\ 
 & (0.421) & (0.367) & (0.177) \\ 
 \rule{0pt}{3ex} 
 Work assignment:Robot & $-$0.303 & $-$0.335 & $-$0.077 \\ 
 & (0.417) & (0.363) & (0.175) \\ 
 \rule{0pt}{3ex} 
 Work assignment:Statistical model & $-$1.048$^{*}$ & $-$1.439$^{**}$ & $-$0.099 \\ 
 & (0.424) & (0.369) & (0.178) \\ 
 
  \rule{0pt}{3ex}\hspace{-2mm}  
  \textbf{Control Variable} & \\ 
   \rule{0pt}{3ex} 
 Affinity for technology & 0.074 & 0.097 & 0.054$^{*}$ \\ 
 & (0.058) & (0.050) & (0.024) \\ 
\hline
\textit{R$^{2}$} & 0.420 & 0.431 & 0.110 \\ 
\textit{F} (\textit{df} = 12; 520) & 31.364$^{**}$ & 32.765$^{**}$ & 5.332$^{**}$ \\ 
\hline 
\multicolumn{4}{l} {\textit{Note:} The results for the tasks can be interpreted in comparison to the}\\
\multicolumn{4}{l} {task work evaluation. The results for the terms can be interpreted in}\\
\multicolumn{4}{l} { comparison to the term artificial intelligence. The columns Fairness,}\\
 \multicolumn{4}{l}{Trust, Justice show estimates for the coefficients and respective }\\
 \multicolumn{4}{l}{standard errors in brackets.}\\
\multicolumn{4}{l}{*\textit{p} < .05, **\textit{p} < .01. \textit{N} = 533.}
\end{tabular} 
\end{table}

\begin{table}[h]
 \centering
  \caption{Results of the linear regression for the comparison of human manager versus the different terms to refer to ADM systems for the task work evaluation in Study 2.}
  \label{tab:regression_study2_evaluation}
\begin{tabular}{lccc}
\hline 
\\[-2.8ex] & Fairness \phantom{te} & Trust & Justice \\ 
\hline \\[-1.8ex] 
 Constant & 5.765$^{**}$ & 4.589$^{**}$ & 2.686$^{**}$ \\ 
 & (0.229) & (0.192) & (0.100) \\ 
 \rule{0pt}{3ex} 
 Artificial intelligence & $-$2.692$^{**}$ & $-$2.055$^{**}$ & $-$0.431$^{**}$ \\ 
 & (0.333) & (0.279) & (0.146) \\ 
 \rule{0pt}{3ex} 
 Algorithm & $-$2.037$^{**}$ & $-$1.538$^{**}$ & $-$0.360$^{*}$ \\ 
 & (0.318) & (0.267) & (0.139) \\ 
 \rule{0pt}{3ex} 
 Automated system & $-$2.391$^{**}$ & $-$1.845$^{**}$ & $-$0.376$^{**}$ \\ 
 & (0.324) & (0.271) & (0.142) \\ 
 \rule{0pt}{3ex} 
 Computer program & $-$2.518$^{**}$ & $-$1.849$^{**}$ & $-$0.307$^{*}$ \\ 
 & (0.332) & (0.278) & (0.145) \\ 
 \rule{0pt}{3ex} 
 Robot & $-$2.447$^{**}$ & $-$1.594$^{**}$ & $-$0.355$^{*}$ \\ 
 & (0.317) & (0.266) & (0.139) \\ 
 \rule{0pt}{3ex} 
 Statistical model & $-$1.820$^{**}$ & $-$1.005$^{**}$ & $-$0.466$^{**}$ \\ 
 & (0.328) & (0.275) & (0.144) \\ 
 
   \rule{0pt}{3ex}\hspace{-2mm}  
  \textbf{Control Variable} & \\ 
   \rule{0pt}{3ex} 
 Affinity for technology & 0.277$^{**}$ & 0.224$^{**}$ & 0.141$^{**}$ \\
 & (0.082) & (0.069) & (0.036) \\ 
\hline
\textit{R$^{2}$} & 0.277 & 0.246 & 0.091 \\ 
\textit{F} (\textit{df} = 7; 298) & 16.288$^{**}$ & 13.904$^{**}$ & 4.276$^{**}$ \\ 
\hline 
  \multicolumn{4}{l}{\textit{Note:} The results for the terms can be interpreted in comparison}\\
  \multicolumn{4}{l}{to the human manager. The columns Fairness, Trust, Justice show }\\
  \multicolumn{4}{l}{estimates and respective standard errors in brackets.}\\
  \multicolumn{4}{l}{*\textit{p} < .05, **\textit{p} < .01. \textit{N} = 306.}
\end{tabular} 
\end{table}

\begin{table}[h]
 \centering
  \caption{Results of the linear regression for the comparison of human manager versus the different terms to refer to ADM systems for the task work assignment in Study 2.}
  \label{tab:regression_study2_assignment}
\begin{tabular}{lccc}
\hline 
\\[-2.8ex] & Fairness \phantom{te} & Trust & Justice \\ 
\hline \\[-1.8ex] 
 Constant & 5.958$^{**}$ & 5.378$^{**}$ & 2.686$^{**}$ \\ 
 & (0.168) & (0.160) & (0.077) \\ 
 \rule{0pt}{3ex} 
 Artificial intelligence & $-$0.342 & $-$0.441$^{*}$ & $-$0.072 \\  & (0.233) & (0.222) & (0.107) \\ 
 \rule{0pt}{3ex} 
 Algorithm & $-$0.212 & $-$0.644$^{**}$ & 0.117 \\ 
 & (0.237) & (0.226) & (0.109) \\ 
 \rule{0pt}{3ex} 
 Automated system & $-$0.071 & $-$0.514$^{*}$ & $-$0.049 \\ 
 & (0.238) & (0.227) & (0.110) \\ 
 \rule{0pt}{3ex} 
 Computer program & $-$0.046 & $-$0.095 & 0.082 \\ 
 & (0.236) & (0.224) & (0.108) \\ 
 \rule{0pt}{3ex} 
 Robot & $-$0.273 & $-$0.238 & $-$0.025 \\ 
 & (0.240) & (0.228) & (0.110) \\ 
 \rule{0pt}{3ex} 
 Statistical model & $-$0.450 & $-$0.779$^{**}$ & $-$0.164 \\ 
 & (0.238) & (0.227) & (0.110) \\ 
 
    \rule{0pt}{3ex}\hspace{-2mm}  
  \textbf{Control Variable} & \\ 
   \rule{0pt}{3ex} 
 Affinity for technology & $-$0.124$^{*}$ & $-$0.003 & 0.010 \\ 
 & (0.060) & (0.057) & (0.028) \\ 
\hline
\textit{R$^{2}$} & 0.030 & 0.060 & 0.030 \\ 
\textit{F} (\textit{df} = 7; 308) & 1.359 & 2.784$^{**}$ & 1.353 \\ 
\hline 
  \multicolumn{4}{l}{\textit{Note:} The results for the terms can be interpreted in comparison}\\
  \multicolumn{4}{l}{to the human manager. The columns Fairness, Trust, Justice show }\\
  \multicolumn{4}{l}{estimates and respective standard errors in brackets.}\\
  \multicolumn{4}{l}{*\textit{p} < .05, **\textit{p} < .01. \textit{N} = 316.}
\end{tabular} 
\end{table} 

\clearpage

\begin{figure*}[h]
  \centering
  \includegraphics[width=0.99\columnwidth]{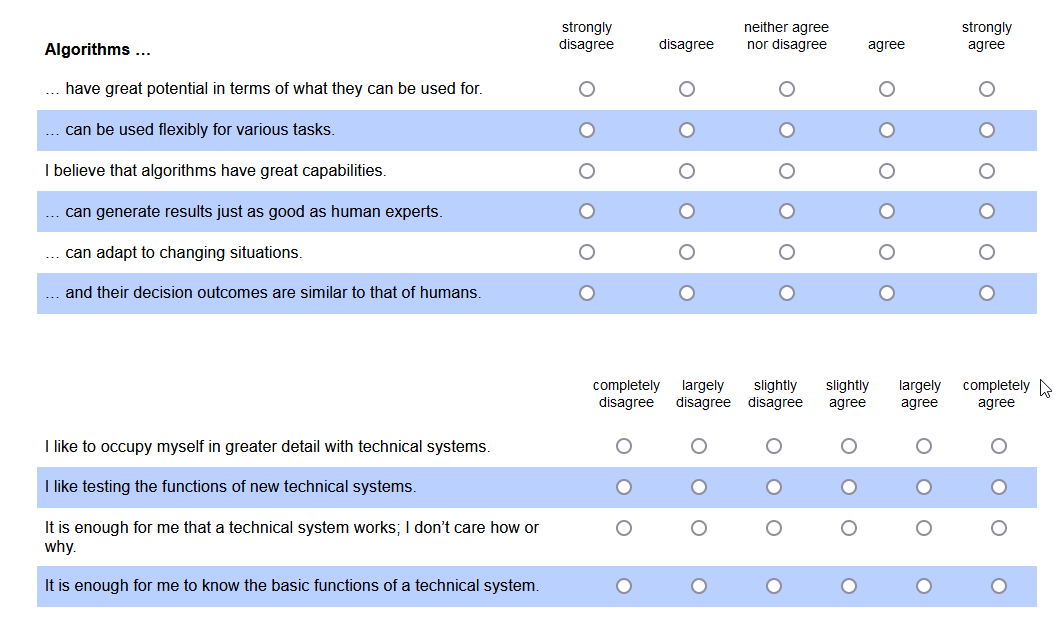}
  \caption{Screenshot from Study 1 where participants reported their perceptions regarding the different terms, here ``the term'' was algorithm.}
  \label{fig:appendix_study1_perceptions}
  \Description{This figure shows how participants experienced the part of Study 1 where they indicated their perceptions of the terms. It shows examples for several items to respond to that assess participants' perceptions of one of the terms.}
\end{figure*}

\begin{figure*}[h]
  \centering
  \includegraphics[width=0.99\columnwidth]{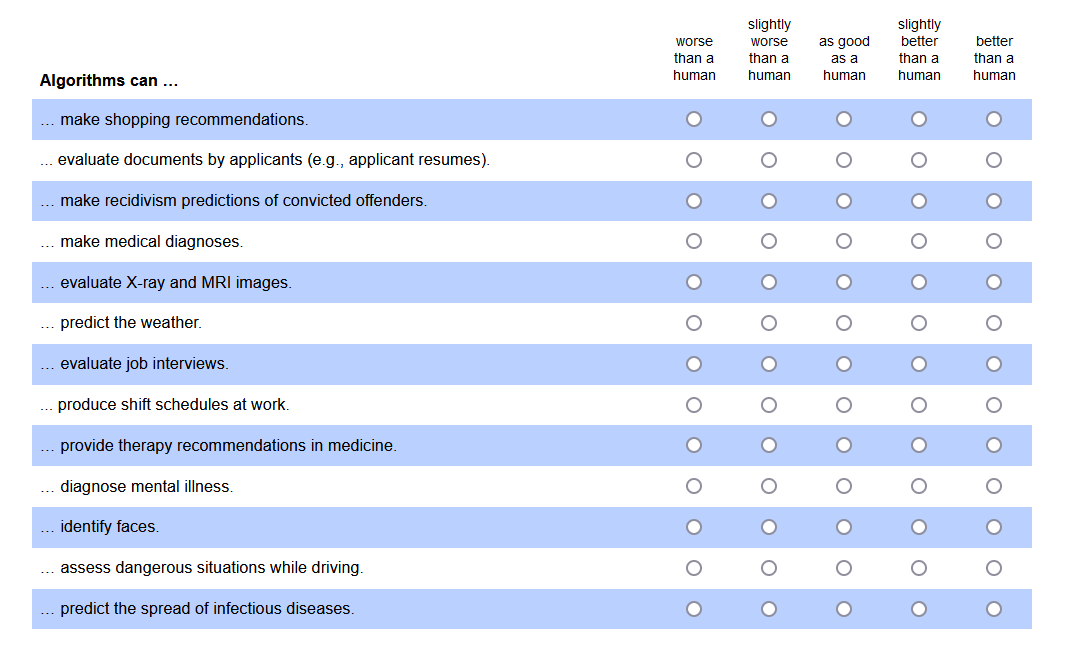}
  \caption{Screenshot from Study 1 where participants reported their evaluation regarding the performance of a respective term in comparison to a human, here ``the term'' was algorithm.}
  \label{fig:appendix_study1_betterthanhuman}
  \Description{This figure shows how participants experienced the part of Study 1 where they indicated whether humans or systems described with a respective term are better able to conduct different tasks. It shows the items that were assessed for all the thirteen tasks that have also been described in the text.}
\end{figure*}

\begin{figure*}[h]
  \centering
  \includegraphics[width=0.99\columnwidth]{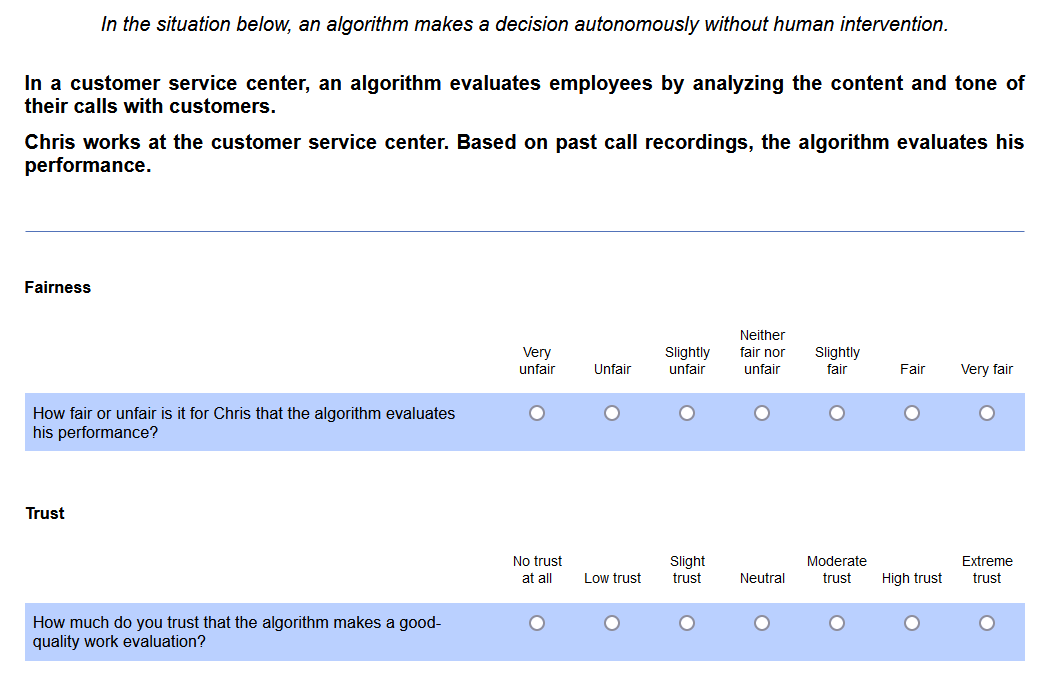}
  \caption{Screenshot from Study 2 where participants reported their evaluation of a respective term regarding the task work evaluation, here ``the term'' was algorithm.}
  \label{fig:appendix_study2_workevaluation}
    \Description{This figure shows how participants experienced Study 2. It shows the textual vignette for the work evaluation task and the items for fairness and trust for this task.}
\end{figure*}

\begin{figure*}[h]
  \centering
  \includegraphics[width=0.99\columnwidth]{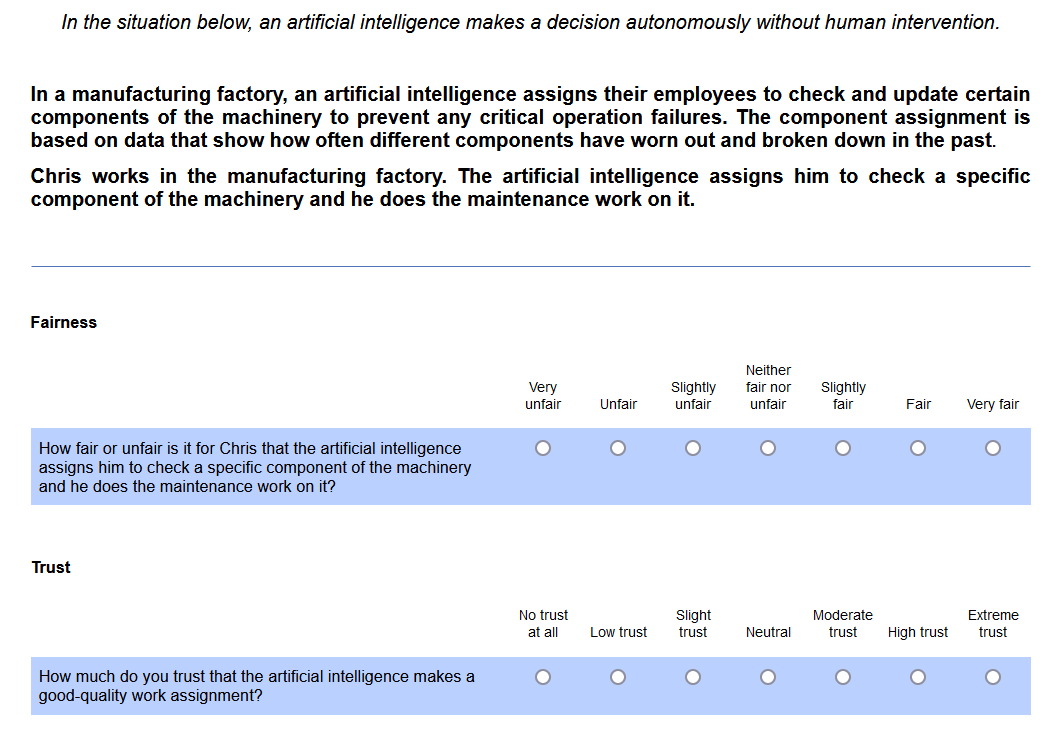}
  \caption{Screenshot from Study 2 where participants reported their evaluation of a respective term regarding the task work assignment, here ``the term'' was artificial intelligence.}
  \label{fig:appendix_study2_workassignment}
    \Description{This figure shows how participants experienced Study 2. It shows the textual vignette for the work assignment task and the items for fairness and trust for this task.}
\end{figure*}

\end{document}